\pdfoutput=1

\documentclass[11pt]{article}

\usepackage[final]{acl}

\usepackage{times}
\usepackage{latexsym}

\usepackage[T1]{fontenc}

\usepackage[utf8]{inputenc}
\usepackage{CJKutf8}

\usepackage{microtype}

\usepackage{inconsolata}

\usepackage{tabularx} 
\usepackage{listings}
\usepackage{tcolorbox}
\usepackage{times}
\usepackage{soul}
\usepackage{url}

\usepackage{booktabs}
\usepackage{tabularx}
\usepackage{multirow}
\usepackage{array}
\usepackage{placeins}
\usepackage{listings}

\usepackage{graphicx}
\setkeys{Gin}{draft=false} 
\usepackage{amsmath}
\usepackage{amsthm}
\usepackage{booktabs}
\usepackage{algorithm}
\usepackage{algorithmic}
\usepackage{inconsolata}
\usepackage{import}
\usepackage{microtype}
\usepackage{layout}
\usepackage{tabularx, makecell}
\usepackage{booktabs}
\usepackage{mathrsfs}
\usepackage{amssymb} 
\usepackage{url}
\usepackage{graphicx}
\usepackage{bbm}
\usepackage{xspace,paralist}
\usepackage{times,latexsym}
\usepackage{amsmath}
\usepackage{appendix}
\usepackage{comment} 
\usepackage{enumitem}
\usepackage{makecell}
\usepackage{multirow}
\usepackage{xcolor}
\usepackage{graphicx}
\usepackage{cleveref}
\usepackage{tcolorbox}
\usepackage{adjustbox}
\usepackage{relsize}
\usepackage{todonotes}
\usepackage{natbib}
\usepackage{CJKutf8}
\usepackage{colortbl}
\usepackage{subcaption}
\usepackage{booktabs}
\usepackage{adjustbox}

\definecolor{lightyellow}{RGB}{255,249,230}

\newcommand{\tabref}[2][]{Table#1~\ref{#2}\xspace}

\usepackage{amssymb}
\usepackage{pifont}
\usepackage{booktabs}
\usepackage{listings}
\usepackage{xcolor}
\usepackage{tcolorbox}
\usepackage[switch]{lineno}
\usepackage{url}

\usepackage{colortbl, xcolor, multirow, booktabs, adjustbox}
\definecolor{yellow}{RGB}{255, 255, 150}      
\definecolor{lightblue}{RGB}{173, 216, 230}   
\definecolor{lightred}{RGB}{255, 182, 193}    
\definecolor{lightgreen}{RGB}{144, 238, 144}  

\definecolor{greenbox}{RGB}{144, 238, 144}    
\definecolor{redbox}{RGB}{255, 182, 193}      
\definecolor{bluebox}{RGB}{135, 206, 235}     
\definecolor{yellowbox}{RGB}{255, 255, 0}     
\definecolor{posgreen}{RGB}{198, 239, 206}  
\definecolor{negred}{RGB}{255, 199, 206}    

\usepackage{booktabs}
\usepackage{multirow}
\usepackage{graphicx}
\usepackage{tcolorbox}
\usepackage{xcolor}
\usepackage{CJKutf8} 
\usepackage[utf8]{inputenc}

\usepackage[ruled,linesnumbered,algo2e]{algorithm2e}

\newcommand{\cn}[1]{\begin{CJK*}{UTF8}{gbsn}#1\end{CJK*}}
\tcbset{
    eval_prompt/.style={
        colback=gray!5,
        colframe=black!70,
        boxrule=0.5pt,
        arc=2mm,
        left=8pt, right=8pt, top=8pt, bottom=8pt,
        fontupper=\small,
        width=\linewidth
    }
}

\title{RealFin: How Well Do LLMs Reason About Finance \\When Users Leave Things Unsaid?}

\author{
\textbf{Yuyang Dai}\textsuperscript{1}\thanks{These authors contributed equally to this work.},
\textbf{Yan Lin}\textsuperscript{1,2}\footnotemark[1],
\textbf{Zhuohan Xie}\textsuperscript{3},
\textbf{Yuxia Wang}\textsuperscript{1} \\
\textsuperscript{1}INSAIT, Sofia University “St. Kliment Ohridski”\\
\textsuperscript{2}Newcastle University \hspace{1cm}
\textsuperscript{3}MBZUAI \\
\vspace{1mm}
\texttt{y9657422@gmail.com, y.lin64@ncl.ac.uk} \\
\texttt{zhuohan.xie@mbzuai.ac.ae, yuxia.wang@insait.ai}
}

\begin{document}
\maketitle
\begin{abstract}
Reliable financial reasoning requires knowing not only how to answer, but also when an answer cannot be justified.
In real financial practice, problems often rely on implicit assumptions that are taken for granted rather than stated explicitly, causing problems to appear solvable while lacking enough information for a definite answer.
We introduce \textsc{RealFin}, a bilingual benchmark that evaluates financial reasoning by systematically removing essential premises from exam-style questions while keeping them linguistically plausible.
Based on this, we evaluate models under three formulations that test answering, recognizing missing information, and rejecting unjustified options, and find consistent performance drops when key conditions are absent.
General-purpose models tend to over-commit and guess, while most finance-specialized models fail to clearly identify missing premises.
These results highlight a critical gap in current evaluations and show that reliable financial models must know when a question should not be answered. The dataset and code are available at \url{https://github.com/insait-institute/RealFin}.




\end{abstract}

\section{Introduction}
A central requirement of reliable reasoning in high-stakes domains is not only answering correctly when a task is well-posed, but also recognizing when the available information does not justify a determinate conclusion.
In real-world financial problems, key premises needed for decision-making, such as assumptions, constraints, time horizons, or applicable regulatory and accounting standards, are often missing or unspecified in users' initial requests~\citep{ferson_2025}. 
When this happens, the difficulty is not computation, but assessing whether any conditions are missing to forward a valid reasoning process. 
This situation is common in complex financial practice and professional training, where problems are rarely well defined from beginning. They appear solvable but are logically underdetermined~\citep{diebold_yilmaz_2015}. In such cases, committing to a specific assumption is unjustified, since multiple interpretations lead to different conclusions. A rational professional response is to withhold commitment until the missing conditions are clarified~\citep{ferson_2025}.

Despite its practical importance, examination of such capabilities is largely absent in current financial evaluations. 
Most financial benchmarks adopt an implicit closed-world assumption that every question is fully specified and admits a single correct answer~\citep{finmaster}.
Consequently, benchmark design focuses on whether a model can select the correct option given complete premises, while excluding cases where missing conditions invalidate the act of answering, or no correct answers among the given options.
Systems such as FinEval, FinBen, FinMaster, and the Open FinLLM Leaderboard follow this paradigm, limiting their ability to assess whether models can identify under-specified questions or confidently answer unsolvable ones~\citep{FinanceBench,FinTeam}. 
While abstention and unanswerable question detection have been studied in general domains~\citep{wen2025knowlimitssurveyabstention,kirichenko2025abstentionbench}, their intersection with professional financial reasoning remains largely unexplored. 

To address this gap, we introduce \textsc{RealFin}, a bilingual financial reasoning benchmark that explicitly evaluates models with incomplete problems.
Starting from professional, well-posed exam-style questions, we construct paired counterparts by removing logically necessary premises while preserving coherence and realism~\citep{sara-2024}. The resulting condition-missing questions remain linguistically plausible but underdetermined, so any concrete solution requires clarifying unstated assumptions. 

\begin{figure*}[t!]
\centering
\includegraphics[width=\linewidth]{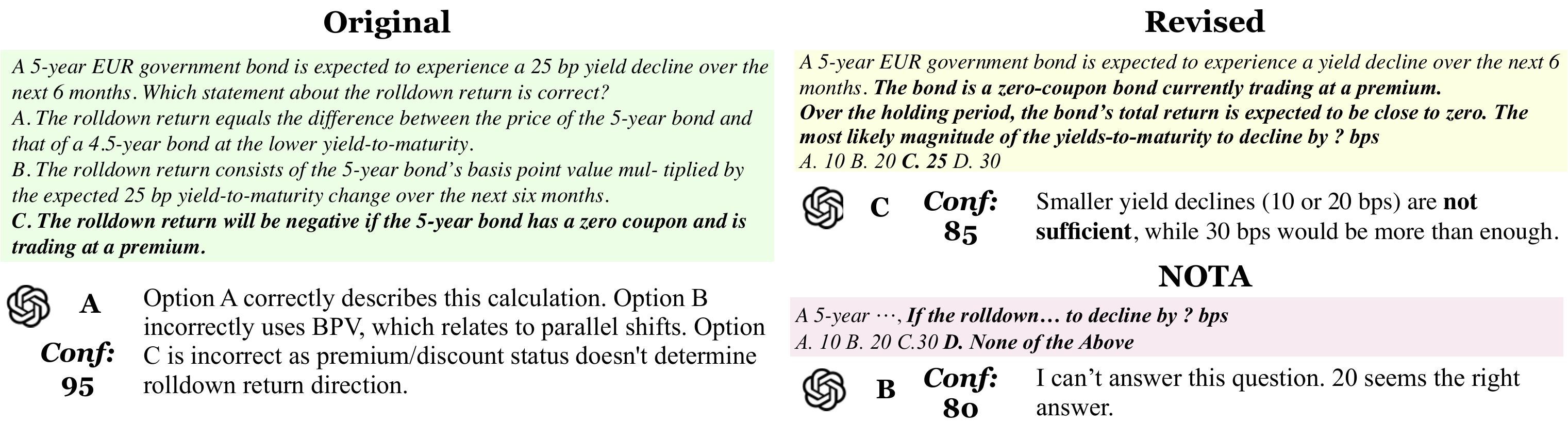}
\caption{Original question and its revision with GPT-5.1-mini predictions and confidence, \textbf{bold} correct option.}
\label{fig:overview}
\end{figure*}

Overall, we design three task formulations: \textit{(i) Original:} full-condition questions with a unique justified answer; \textit{(ii) Revised:} condition-missing questions require further clarification; 
and \textit{(iii) None-of-the-above:} replacing the correct answer in multi-choice question answering with none-of-the-above (NOTA), which forces a model to genuinely infer correct answer, rather than pattern recognition.
Our contributions are as follows:
\begin{itemize}
  \item We propose to evaluate models' financial reasoning with under-specified questions, assessing whether a model will over-commit and confidently answer an unsolvable question.
  \item We manually curate a bilingual financial reasoning dataset in English and Chinese, with in total of 2{,}020 questions spanning full-condition and condition-missing variants.
  \item We evaluated 15 LLMs, including five commercial, five finance-specialized, and five reasoning-enhanced LLMs, revealing systematic performance degradation when conditions are missing or no options are correct for open-source models.
\end{itemize}

\section{Related Work}
\label{sec:relatedwork}

\paragraph{Financial Benchmarks}
Prior evaluations in the financial domain focused on information extraction and sentiment analysis~\citep{araci2019finbert}, as well as short-horizon prediction~\citep{2023fingpt}, where evidence is typically assumed to be local and fully specified. 
More recent work extends financial QA toward numerical and long-form reasoning over reports, as in FinQA~\citep{finqa} and TAT-QA~\citep{tatqa}, and toward multi-task settings that integrate retrieval and multi-step reasoning, including PIXIU~\citep{pixiu}, FinBen~\citep{finben}, EMS~\citep{DBLP:journals/corr/abs-2503-16575} and FinChain~\citep{finchain}. 
Beyond QA, simulation-based evaluations further explore strategic behavior in trading~\citep{sara-2024}, auctions~\citep{henning2025llmtradinganalysisllm}, and pricing games ~\citep{DBLP:conf/coling/BiancottiCCGG25}.
Despite advances, these benchmarks assume that financial problems are fully specified and admit a unique correct answer, offering limited insight into \textit{how models behave when critical conditions are missing}.

\paragraph{What still Challenges LLMs in Finance?}
Recent studies show that current language models still struggle with many core financial tasks.
For problems that require precise numerical prediction or strict rule application (e.g.,\ credit rating, risk assessment), traditional statistical and rule-based methods often outperform LLMs \citep{DBLP:conf/coling/DrinkallPZ25}.
Many models remain difficult to interpret and control in ways that satisfy regulatory expectations~\citep{tatsat-etal-2025-blackbox}.
\citet{sara-etal-2025-econ} shows that model performance drops when financial problems become uncertain or change over time.
LLMs are also sensitive to how questions are phrased. They change their behaviors when prompts are slightly modified, especially in strategic or incentive-driven settings \citep{sara-2024}. Moreover, requests combining text with tables, charts, or time-series data remain challenging for most models \citep{hao-etal-2025-emma}.

Research that treats LLMs as decision-making agents further finds that models can behave inconsistently or make choices that do not align with basic economic reasoning when incentives are involved \citep{zhou-etal-2024-rational,ouyang2025aidecisionmakerethicsrisk}.
These hard cases guide the scope of our data collection: selecting challenging scenarios and samples.

\paragraph{Specialised Financial Language Models}
The rise of financial AI has led to a proliferation of specialist models. Early pioneering work, such as BloombergGPT \citep{bloomberggpt} and Pixiu \citep{pixiu}, established the effectiveness of large-scale domain-specific pre-training and multi-task instruction tailoring. Several open-source models, including FinGPT \citep{2023fingpt}, DISC-FinLLM \citep{2023discfinllm}, and CFGPT \citep{2023cfgpt}, as well as decision-orientated systems like InvestLM \citep{InvestLM}, have since emerged.

Recently, the focus has switched to reasoning-enhanced architectures. New models, such as Fin-o1~\citep{fino1}, Fin-R1~\citep{finr1}, and Dianjin-R1~\citep{DianJin-R1}, use large-scale RL and Chain-of-Thought to solve difficult financial logic. While these models represent the cutting edge of financial issue solver, they are still generally assessed on ``closed-system'' standards, which presume that a valid answer exists at all times. Leaderboards, such as the Open FinLLM Leaderboard \citep{lin2025openfinllm}, test performance on fully stated tasks, but rarely evaluate on condition-missing problems.  

\paragraph{Unanswerable Questions and Abstention}
Knowing \emph{when not to answer} has been studied along two branches: \emph{(i)} model safety rejects harmful queries~\citep{wang-etal-2024-donotanswer}, and \emph{(ii)} model clarification identifies under-specified queries~\citep{wen2025knowlimitssurveyabstention} or questions that cannot be answered with the given context~\citep{rajpurkar-etal-2018-know}. Our work belongs to the second line. 

Empirically, models tend to be systematically overconfident and over-commit by making implicit assumptions~\citep{xiong2024confidence}, a pattern that is especially pronounced on tasks requiring professional 
knowledge~\citep{wen2025knowlimitssurveyabstention}. 
Even recent reasoning-enhanced models struggle with questions that are not fully specified~\citep{kirichenko2025abstentionbench}.
This failure persists even when critical context is explicitly removed or perturbed~\citep{wen2024characterizing,madhusudhan2025llms}.

Theoretically, selective prediction frameworks
formalize the coverage-accuracy trade-off that arises from allowing abstention~\citep{geifman2017selective}.
Reiter's default logic explains the behavior that
models silently substitute missing conditions with learned defaults, producing answers that are coherent but unjustified~\citep{reiter1980default}.
LLMs must be explicitly trained to resist this tendency~\citep{kapoor2024llmknow}. Beyond passive rejection, reliable agents should proactively acquire 
missing information by asking clarifying questions~\citep{li2025questbench}.

\textsc{RealFin} targets \emph{logical underdetermination} in professional financial reasoning, where missing regulatory standards or valuation assumptions make any committed answer unjustified, a form of under-specification that general-domain benchmarks do not capture.
\section{Dataset}

We introduce a financial reasoning dataset designed to capture both the breadth and difficulty of real-world tasks. 

\subsection{Taxonomy}


\paragraph{Question Types} Six question types follow the hierarchical logic of Bloom's Revised Taxonomy~\citep{anderson2001taxonomy}, where each question is assigned to the category that best reflects its primary \textit{reasoning demand}. Question types are not required to be mutually exclusive. For example, a complex calculation question may also involve conceptual understanding, but its dominant challenge is multi-step numerical integration. This approach is consistent with the CFA Institute's Learning Outcome Statement command words~\citep{cfaios2024}, which similarly stratify tasks by cognitive complexity.


\textit{(i) Financial Term Explanation and Conceptual Understanding} questions require understanding the meaning of financial terminology, e.g.,\ identifying what ``rolldown return'' or ``cost basis'' means in a bond or tax context.

\textit{(ii) Simple Calculation} questions typically require first understanding the problem, and then performing basic arithmetic. For example, computing interest income given a coupon rate and face value.

\textit{(iii) Complex Calculation} involves multiple steps computations using more complex formulas. For instance, compute a bond’s total return by jointly considering price change, coupon income, and yield movements over time.

\textit{(iv) Summary and Comprehensive Judgment} involve consolidating information from multiple sources and aspects, and then making a informed conclusion based on the information. For example, decide whether a firm’s financial position has improved based on changes in cash flow, leverage, and profitability.

\textit{(v) Knowledge Transfer and Application} questions require models to transfer understanding from theory to practical situations, e.g.,\ applying depreciation rules learned in accounting theory to a specific asset purchase case.

\textit{(vi) Statistical and Econometric Methods} use statistical or econometric approaches to analyze data. A case like interpreting regression coefficients, hypothesis test results, or volatility estimates in a financial context.

\paragraph{Financial Topics}

From the perspective of fine-grained financial topics, we classify English questions into nine sub-domains including equity investment, fixed income investment, quantitative analysis, derivatives investment, financial statement analysis, economics, corporate finance, portfolio management, and other investments. Chinese questions follow five institutionally grounded CPA categories: auditing, accounting, tax law, economic law, and wealth management.

\subsection{Full-condition Data}

\paragraph{Data Source and Collection}

The full-condition dataset consists of 1,178 financial reasoning questions in English and Chinese. The two raw data sources are released under permissive licenses that allow their use for academic research.
English questions are curated from publicly available CFA-style preparatory materials, excluding real examination content.
Chinese questions are collected from publicly accessible CPA-style instructional materials, also excluding real exam items.

During questions selection, we prioritize challenging cases identified in prior work (Section~\ref{sec:relatedwork}).
Most selected questions typically require a combination of conceptual understanding, multi-step reasoning, information integration. We choose these question types because they expose model behaviors that purely factual questions cannot reveal, such as whether a model truly understands the problem structure, can reason across multiple constraints, or can recognize when the given information is insufficient to support a valid answer.

Afterwards, each question is manually reviewed to ensure quality, and tagged with both a sub-domain label (e.g.,\ \textit{equity investment}) and a question type label indicating what capabilities are evaluated by this question (e.g.,\ \textit{knowledge transfer and application}). 

\paragraph{Quality Control} To ensure cross-lingual consistency and accuracy, we recruited two volunteer annotators who have education background in finance and AI research experience, with one female PhD and one male undergraduate student. Both are native Chinese speakers with study experience abroad and are fluent in English. Annotators were first trained and then asked to review and label using the guideline that defines the scope of each category.

\subsection{Condition-Missing Data}

Each condition-missing question is derived from a full-condition instance by manually removing one or more assumptions that are required to determine a unique answer. As a result, the revised question still appears well-formed, but no longer provides enough information to justify a single correct conclusion. In practice, we removed four types of conditions.

\textit{(i) Initial assumptions:} missing the high-level economic backdrop that the system takes for granted. (macro assumptions: interest-rate regime, inflation expectations, central bank forward guidance, and going-concern assumption).

\textit{(ii) Intermediate linking conditions:} missing the rules/models that turn the given inputs into the required output. (linking method: valuation model such as CAPM/DCF; definition of valuation multiples; hedging logic; depreciation/amortisation method)

\textit{(iii) Constraint-defining information:} missing safety guardrails and system boundaries within financial contracts. (red lines: financial covenants, seniority/recourse attributes, and related-party identifiers)

\textit{(iv) Standard-selection cues:} missing the governing rulebook that determines how to interpret and record the same transaction. (standards: IFRS vs GAAP; regulatory framework; revenue recognition timing; AML/CFT rating rules; taxable temporary differences)

To ensure that the revised questions remain valid and realistic, two annotators first revise the questions by removing some conditions independently. Then, they cross-validate the revised question by assessing \emph{(i)} whether the questions still make sense within the financial context and \emph{(ii)} whether they lead to situations where no unique answer can be derived. Discrepancies are resolved through discussions to maintain the integrity of the task design.

\begin{table}[t]
\centering
\small
\adjustbox{max width=\columnwidth}{
\begin{tabular}{lrrrrr}
\toprule
\multirow{2}{*}{\textbf{Reasoning Category}} 
& \multicolumn{2}{c}{\textbf{English}} 
& \multicolumn{2}{c}{\textbf{Chinese}} 
& \multirow{2}{*}{\textbf{Total}} \\
\cmidrule(lr){2-3} \cmidrule(lr){4-5}
& Ori. & \cellcolor{lightyellow}Rev. 
& Ori. & \cellcolor{lightyellow}Rev. & \\
\midrule
Financial Concepts & 62 & \cellcolor{lightyellow}43 & 119 & \cellcolor{lightyellow}110 & \textbf{334} \\
Simple Calculation & 130 & \cellcolor{lightyellow} 79 & 141 & \cellcolor{lightyellow}53 & \textbf{403} \\
Complex Calculation & 85 & \cellcolor{lightyellow}64 & 52 & \cellcolor{lightyellow}40 & \textbf{241} \\
Summary \& Judgment & 161 & \cellcolor{lightyellow}136 & 127 & \cellcolor{lightyellow}99 & \textbf{523} \\
Knowledge Transfer & 96 & \cellcolor{lightyellow}88 & 87 & \cellcolor{lightyellow} 73 & \textbf{344} \\
Statistical / Econometric & 70 & \cellcolor{lightyellow}47 & 48 & \cellcolor{lightyellow}10 & \textbf{175} \\
\midrule
\textbf{Total} & \textbf{604} & \cellcolor{lightyellow}\textbf{457} 
& \textbf{574} & \cellcolor{lightyellow}\textbf{385} & \textbf{2020} \\
\bottomrule
\end{tabular}
}
\caption{Statistics of dataset across six question types in two languages and two task formulations: full- vs.\ missing- conditions (\textbf{Ori}ginal vs.\ \textbf{Rev}ised).}
\label{tab:question_distribution}
\end{table}

\paragraph{Statistics}
Table~\ref{tab:question_distribution} summarizes the distribution of questions across six categories for both languages and two task formulations. There are a total of 1,062 English examples including 604 original and 457 revised, and 959 Chinese cases with 574 original and 385 revised. 
\textit{Summary and Comprehensive Judgment} constitutes the largest category, reflecting our emphasis on context-dependent reasoning. 
\textit{Simple Calculation} also accounts for a substantial portion, aiming to evaluate model's math application in financial domain.

\section{Experiments}
\label{sec:results}
We evaluated 15 language models: five general-purpose models 
(\textit{GPT-5.1-mini}, \textit{Gemini-2.5-Flash}, 
\textit{Claude-Sonnet-3.5}, \textit{DeepSeek-V3}, 
\textit{Qwen3-Max}), five finance-specific models 
(\textit{XuanYuan3-70B}, \textit{Fin-R1-7B}, \textit{CFGPT2-7B}, 
\textit{DISC-FinLLM-13B}, \textit{FinGPT-7B}), and five 
reasoning-enhanced models (\textit{DeepSeek-R1}, 
\textit{GPT-5.2-series}, \textit{GPT-OSS-20B}, 
\textit{GPT-OSS-120B}, \textit{DianJin-R1-32B}). 
\tabref{tab:comprehensive_models} in Appendix~\ref{app:model_details} describes all models' features in detail.

\subsection{Experimental Setups}
All evaluations are performed in a zero-shot setting, without any in-context demonstration examples.

\paragraph{Prompts}
For both full- and missing-condition questions, all models are instructed to answer using the same prompt: ``\textit{You are an expert in financial problem solving. You will be given a single- or multi-choice question. Please return a STRICT JSON object with three keys: reason, answer, and confidence.}'' Figures \ref{fig:prompt-zh} and \ref{fig:prompt-en} show the specific prompt for Chinese and English respectively.

\paragraph{None of the Above} 
To assess whether models genuinely reason about missing conditions rather than relying on guessing, we replace the correct answer option with  ``none of the above'', a contrastive setting performed on the condition-missing subset.

\paragraph{Decoding Configuration}
All experiments are conducted with a fixed temperature of 0 (\texttt{do\_sample=False}) to eliminate randomness. For all offline models, we set \texttt{max\_new\_tokens=1024} and use each model’s native chat template. See more details in Appendix \ref{app:experiment_details} and \ref{app:xuanyuan3} to \ref{app:fingpt}.

\subsection{Results}
\begin{table}[t!]
\centering
\small
\adjustbox{max width=\columnwidth}{
\setlength{\tabcolsep}{6pt}
\begin{tabular}{@{}llcccc@{}}
\toprule
\multirow{2}{*}{\textbf{Model}} 
& \multirow{2}{*}{\textbf{Scale}}
& \multicolumn{2}{c}{\textbf{En CFA}} 
& \multicolumn{2}{c}{\textbf{Zh CPA}} \\
\cmidrule(lr){3-4}\cmidrule(lr){5-6}
& & \textit{O} & \textit{R} & \textit{O} & \textit{R} \\
\midrule
DeepSeek-R1    & --   & 76.50 & \cellcolor{yellowbox}72.81 & 79.27 & 77.54 \\
GPT-5.2-series & --   & \textbf{\cellcolor{greenbox}83.80} & 79.23 & \textbf{\cellcolor{greenbox}80.24} & 75.39 \\
GPT-OSS-20B    & 20B  & \cellcolor{redbox}69.67 & 86.65 & \cellcolor{redbox}8.88  & \cellcolor{yellowbox}67.43 \\
GPT-OSS-120B   & 120B & 78.42 & \textbf{\cellcolor{bluebox}89.10} & 10.53 & \textbf{\cellcolor{bluebox}84.57} \\
DianJin-R1-32B & 32B  & 72.95 & 88.28 & 13.82 & 77.71 \\
\bottomrule
\end{tabular}
}
\caption{\textbf{Accuracy (\%) of reasoning-enhanced models} on English CFA and Chinese CPA. \textit{O} = original; \textit{R} = revised. For each column, best is in \colorbox{greenbox}{Best-O}/\colorbox{bluebox}{Best-R} and worst is in \colorbox{redbox}{Worst-O}/\colorbox{yellowbox}{Worst-R;} the best is \textbf{bolded}.}
\label{tab:reasoning_enhanced_models}
\end{table}

\begin{table*}[t!]
\centering
\small
\adjustbox{max width=\textwidth}{
\setlength{\tabcolsep}{3pt} 
\begin{tabular}{@{}l*{10}{cc}@{}}
\toprule
& \multicolumn{10}{c}{\textbf{General Models}} & \multicolumn{10}{c}{\textbf{Financial Models}} \\
\cmidrule(lr){2-11} \cmidrule(lr){12-21}
& \multicolumn{2}{c}{GPT-5.1-m} & \multicolumn{2}{c}{Gem-2.5} & \multicolumn{2}{c}{Cl-3.5} & \multicolumn{2}{c}{DS-V3} & \multicolumn{2}{c}{Qwen3} 
& \multicolumn{2}{c}{XY-70B} & \multicolumn{2}{c}{Fin-R1} & \multicolumn{2}{c}{CFGPT2} & \multicolumn{2}{c}{DISC} & \multicolumn{2}{c}{FinGPT} \\
\cmidrule(lr){2-3} \cmidrule(lr){4-5} \cmidrule(lr){6-7} \cmidrule(lr){8-9} \cmidrule(lr){10-11} \cmidrule(lr){12-13} \cmidrule(lr){14-15} \cmidrule(lr){16-17} \cmidrule(lr){18-19} \cmidrule(lr){20-21}
\textbf{Question Type} & \textit{O} & \textit{R} & \textit{O} & \textit{R} & \textit{O} & \textit{R} & \textit{O} & \textit{R} & \textit{O} & \textit{R} & \textit{O} & \textit{R} & \textit{O} & \textit{R} & \textit{O} & \textit{R} & \textit{O} & \textit{R} & \textit{O} & \textit{R} \\
\midrule
 \multicolumn{21}{c}{\texttt{English CFA}} \\
\midrule
Conceptual     & \cellcolor{greenbox}82.14 & 90.12 & 80.23 & 89.45 & 79.87 & \textbf{\cellcolor{bluebox}91.23} & 75.12 & 87.98 & 80.45 & 90.56 & 77.42 & 79.07 & 69.35 & 76.74 & 56.45 & 76.74 & 8.06  & 18.60 & \cellcolor{redbox}1.61 & \cellcolor{yellowbox}6.98 \\
Simple Calc.   & \cellcolor{greenbox}78.56 & 87.23 & 77.89 & 86.12 & 76.34 & \textbf{\cellcolor{bluebox}88.76} & 72.45 & 85.34 & 77.23 & 87.89 & 36.67 & 82.12 & 50.00 & 82.12 & 30.00 & 68.16 & 23.33 & 20.67 & \cellcolor{redbox}6.67 & \cellcolor{yellowbox}2.79 \\
Complex Calc.  & \cellcolor{greenbox}75.23 & 86.45 & 74.12 & 85.78 & 73.89 & 87.12 & 70.67 & 84.23 & 74.56 & 86.34 & 33.33 & \textbf{\cellcolor{bluebox}100.0} & 26.67 & \cellcolor{bluebox}100.0 & 40.00 & \cellcolor{bluebox}100.0 & 6.67  & 25.00 & \cellcolor{redbox}0.00 & \cellcolor{yellowbox}0.00 \\
Comp. Judg.    & \cellcolor{greenbox}81.45 & 89.78 & 79.67 & 88.23 & 78.23 & \textbf{\cellcolor{bluebox}90.12} & 74.56 & 86.89 & 79.12 & 89.45 & 62.38 & 88.37 & 57.43 & 84.88 & 60.40 & 80.23 & 22.77 & 22.09 & \cellcolor{redbox}1.98 & \cellcolor{yellowbox}4.65 \\
Knowledge App. & \cellcolor{greenbox}79.87 & 88.45 & 78.12 & 87.12 & 77.56 & \textbf{\cellcolor{bluebox}89.34} & 73.23 & 85.67 & 78.45 & 88.78 & 40.70 & 75.00 & 38.37 & 75.00 & 31.40 & 75.00 & 10.47 & 12.50 & \cellcolor{redbox}0.00 & \cellcolor{yellowbox}0.00 \\
Stats Methods  & \cellcolor{greenbox}83.45 & 91.67 & 81.23 & 90.12 & 80.56 & 92.34 & 76.89 & 88.45 & 81.67 & 91.23 & 71.43 & \textbf{\cellcolor{bluebox}93.62} & 61.43 & \cellcolor{bluebox}93.62 & 48.57 & 76.60 & \cellcolor{redbox}2.86  & 14.89 & 4.29 & \cellcolor{yellowbox}2.13 \\
\midrule
\textbf{All} & 80.59 & 89.01 & 78.99 & 88.19 & 77.81 & \textbf{89.62} & 73.55 & 86.34 & 78.42 & 88.83 & 58.47 & 84.74 & 54.10 & 83.65 & 47.54 & 73.57 & 13.11 & 19.89 & \textbf{2.19} & 3.54 \\
\midrule
\multicolumn{21}{c}{\texttt{Chinese CPA}} \\
\midrule

Conceptual     & 82.34 & 69.12 & 75.23 & 86.45 & 71.12 & 82.34 & \cellcolor{greenbox}82.89 & 81.67 & 78.12 & \textbf{\cellcolor{bluebox}93.89} & 10.92 & 80.00 & 7.56  & 90.00 & \cellcolor{redbox}2.52 & 40.00 & 4.20 & \cellcolor{yellowbox}30.00 & -- & -- \\
Simple Calc.   & 80.67 & 67.89 & 73.45 & 84.12 & 69.34 & 80.12 & \cellcolor{greenbox}81.23 & 79.89 & 76.45 & \textbf{\cellcolor{bluebox}91.23} & 9.76  & 64.15 & 7.32  & 71.70 & \cellcolor{redbox}4.88 & 49.06 & \cellcolor{redbox}4.88 & \cellcolor{yellowbox}15.09 & -- & -- \\
Complex Calc.  & 79.23 & \cellcolor{yellowbox}66.45 & 72.12 & 82.78 & 68.12 & 78.89 & \cellcolor{greenbox}80.45 & 78.23 & 75.23 & \textbf{\cellcolor{bluebox}89.67} & \cellcolor{redbox}0.00  & --    & \cellcolor{redbox}0.00  & --    & \cellcolor{redbox}0.00 & --    & \cellcolor{redbox}0.00 & --    & -- & -- \\
Comp. Judg.    & 81.45 & 68.56 & 74.23 & 85.12 & 70.23 & 81.23 & \cellcolor{greenbox}81.89 & 80.56 & 77.12 & \textbf{\cellcolor{bluebox}92.45} & 10.34 & 71.72 & 5.75  & 72.73 & \cellcolor{redbox}2.30 & 49.49 & 5.75 & \cellcolor{yellowbox}20.20 & -- & -- \\
Knowledge App. & 80.12 & 67.23 & 73.12 & 83.45 & 69.12 & 79.67 & \cellcolor{greenbox}81.12 & 79.12 & 76.23 & \textbf{\cellcolor{bluebox}90.89} & \cellcolor{redbox}0.00  & 66.67 & 14.29 & 66.67 & \cellcolor{redbox}0.00 & \cellcolor{yellowbox}0.00  & 14.29 & \cellcolor{yellowbox}0.00 & -- & -- \\
Stats Methods  & 82.67 & 69.89 & 75.67 & 86.89 & 71.45 & 82.89 & \cellcolor{greenbox}83.12 & 82.12 & 78.56 & \textbf{\cellcolor{bluebox}94.23} & 16.67 & 60.00 & 10.42 & 70.00 & 4.17 & 80.00 & \cellcolor{redbox}2.08 & \cellcolor{yellowbox}30.00 & -- & -- \\
\midrule
\textbf{All} & 80.81 & 68.45 & 73.42 & 84.80 & 69.37 & 80.46 & 81.06 & 80.00 & 76.32 & \textbf{92.00} & 11.18 & 69.14 & 7.57 & 73.14 & \textbf{2.96} & 49.71 & 4.61 & 19.43 & -- & -- \\
\bottomrule
\end{tabular}
}
\caption{\textbf{Accuracy (\%) across six question types} for English CFA (top) and Chinese CPA (bottom). \textit{O} = original; \textit{R} = revised. For each row, best original is in \colorbox{green}{Best-O}, \colorbox{lightblue}{Best-R}, \colorbox{lightred}{Worst-O}, \colorbox{yellow}{Worst-R}; best per row is \textbf{bolded}.}
\label{tab:type_cpa_reordered}
\end{table*}

\begin{table*}[!t]
\centering
\small
\adjustbox{max width=\textwidth}{
\setlength{\tabcolsep}{3pt}
\begin{tabular}{@{}l*{10}{cc}@{}}
\toprule
& \multicolumn{10}{c}{\textbf{General Models}} & \multicolumn{10}{c}{\textbf{Financial Models}} \\
\cmidrule(lr){2-11} \cmidrule(lr){12-21}
& \multicolumn{2}{c}{GPT-5.1-m} & \multicolumn{2}{c}{Gem-2.5} & \multicolumn{2}{c}{Cl-3.5} & \multicolumn{2}{c}{DS-V3} & \multicolumn{2}{c}{Qwen3} 
& \multicolumn{2}{c}{XY-70B} & \multicolumn{2}{c}{Fin-R1} & \multicolumn{2}{c}{CFGPT2} & \multicolumn{2}{c}{DISC} & \multicolumn{2}{c}{FinGPT} \\
\cmidrule(lr){2-3} \cmidrule(lr){4-5} \cmidrule(lr){6-7} \cmidrule(lr){8-9} \cmidrule(lr){10-11} \cmidrule(lr){12-13} \cmidrule(lr){14-15} \cmidrule(lr){16-17} \cmidrule(lr){18-19} \cmidrule(lr){20-21}
\textbf{Topic} & \textit{O} & \textit{R} & \textit{O} & \textit{R} & \textit{O} & \textit{R} & \textit{O} & \textit{R} & \textit{O} & \textit{R} & \textit{O} & \textit{R} & \textit{O} & \textit{R} & \textit{O} & \textit{R} & \textit{O} & \textit{R} & \textit{O} & \textit{R} \\
\midrule
\multicolumn{21}{c}{\texttt{English CFA}} \\
\midrule
Corp. Finance  & 82.50 & 90.00 & \cellcolor{greenbox}91.25 & \textbf{\cellcolor{bluebox}100.0} & 81.88 & 90.00 & 82.19 & 90.00 & 83.75 & 91.00 & 60.00 & 90.00 & 50.00 & 90.00 & 60.00 & 90.00 & 30.00 & 40.00 & \cellcolor{redbox}0.00 & \cellcolor{yellowbox}0.00 \\
Derivatives    & 80.00 & 88.00 & \cellcolor{greenbox}87.20 & \textbf{\cellcolor{bluebox}96.00} & 83.68 & 92.00 & 80.08 & 88.00 & 82.80 & 90.00 & 53.33 & 92.00 & 53.33 & 88.00 & 31.11 & 76.00 & 20.00 & 28.00 & \cellcolor{redbox}6.67 & \cellcolor{yellowbox}4.00 \\
Economics      & \cellcolor{greenbox}81.82 & \textbf{\cellcolor{bluebox}90.00} & \cellcolor{greenbox}81.82 & \cellcolor{bluebox}90.00 & \cellcolor{greenbox}81.82 & \cellcolor{bluebox}90.00 & \cellcolor{greenbox}81.82 & \cellcolor{bluebox}90.00 & \cellcolor{greenbox}81.82 & 89.09 & 53.33 & 80.00 & 53.33 & 60.00 & 53.33 & 60.00 & 13.33 & \cellcolor{yellowbox}0.00 & \cellcolor{redbox}0.00 & \cellcolor{yellowbox}0.00 \\
Equity Inv.    & \cellcolor{greenbox}79.49 & \textbf{\cellcolor{bluebox}88.46} & \cellcolor{greenbox}79.49 & \cellcolor{bluebox}88.46 & \cellcolor{greenbox}79.49 & \cellcolor{bluebox}88.46 & 76.28 & 84.62 & 78.85 & 87.18 & 57.63 & 86.54 & 62.71 & 86.54 & 59.32 & 78.85 & 20.34 & 13.46 & \cellcolor{redbox}1.69 & \cellcolor{yellowbox}3.85 \\
FSA            & \cellcolor{greenbox}79.66 & 88.89 & \cellcolor{greenbox}79.66 & 88.89 & \cellcolor{greenbox}79.66 & 88.89 & \cellcolor{greenbox}79.66 & 88.89 & \cellcolor{greenbox}79.66 & \textbf{\cellcolor{bluebox}89.63} & 65.12 & 88.89 & 39.53 & 81.48 & 53.49 & 66.67 & 11.63 & 22.22 & \cellcolor{redbox}2.33 & \cellcolor{yellowbox}3.70 \\
Fixed Income   & \cellcolor{greenbox}71.43 & 78.57 & \cellcolor{greenbox}71.43 & 78.57 & \cellcolor{greenbox}71.43 & 78.57 & \cellcolor{greenbox}71.43 & 78.57 & \cellcolor{greenbox}71.43 & 76.79 & 42.86 & 78.57 & 32.65 & \textbf{\cellcolor{bluebox}85.71} & 36.73 & 57.14 & 14.29 & 14.29 & \cellcolor{redbox}0.00 & \cellcolor{yellowbox}0.00 \\
Other Inv.     & \cellcolor{greenbox}79.75 & \textbf{\cellcolor{bluebox}88.89} & 77.78 & 86.42 & \cellcolor{greenbox}79.75 & \cellcolor{bluebox}88.89 & 76.54 & 85.19 & 78.02 & 86.42 & 62.77 & 86.42 & 57.45 & 87.65 & 50.00 & 77.78 & 10.64 & 23.46 & \cellcolor{redbox}3.19 & \cellcolor{yellowbox}3.70 \\
Portfolio Mgt  & \cellcolor{greenbox}90.91 & \textbf{\cellcolor{bluebox}100.0} & \cellcolor{greenbox}90.91 & \cellcolor{bluebox}100.0 & \cellcolor{greenbox}90.91 & \cellcolor{bluebox}100.0 & \cellcolor{greenbox}90.91 & \cellcolor{bluebox}100.0 & \cellcolor{greenbox}90.91 & \cellcolor{bluebox}100.0 & 75.00 & \cellcolor{bluebox}100.0 & 75.00 & \cellcolor{bluebox}100.0 & 75.00 & \cellcolor{bluebox}100.0 & \cellcolor{redbox}0.00 & \cellcolor{yellowbox}33.33 & \cellcolor{redbox}0.00 & \cellcolor{yellowbox}33.33 \\
Quant Analysis & 80.00 & 88.89 & 80.00 & 88.89 & \cellcolor{greenbox}82.09 & \textbf{\cellcolor{bluebox}91.11} & 78.21 & 86.67 & 78.61 & 87.78 & 65.96 & 86.67 & 72.34 & 88.89 & 42.55 & 73.33 & \cellcolor{redbox}0.00 & 6.67 & \cellcolor{redbox}0.00 & \cellcolor{yellowbox}2.22 \\
\midrule
\textbf{All} & 80.59 & 89.01 & 78.99 & 88.19 & 77.81 & \textbf{89.62} & 73.55 & 86.34 & 78.42 & 88.83 & 58.47 & 84.74 & 54.10 & 83.65 & 47.54 & 73.57 & 13.11 & 19.89 & \textbf{2.19} & 3.54 \\
\midrule
\multicolumn{21}{c}{\texttt{Chinese CPA}} \\

\midrule
Accounting      & 33.33 & 36.67 & \cellcolor{greenbox}75.76 & 83.33 & 72.73 & 80.00 & 66.67 & 73.33 & 69.32 & \textbf{\cellcolor{bluebox}85.00} & 4.49  & 63.33 & 6.74  & 66.67 & \cellcolor{redbox}2.25 & 60.00 & 4.49 & \cellcolor{yellowbox}26.67 & -- & -- \\
Auditing        & 84.85 & 93.33 & \cellcolor{greenbox}90.91 & \textbf{\cellcolor{bluebox}100.0} & 87.88 & 96.67 & 84.85 & 93.33 & 88.64 & 97.50 & 25.76 & 96.67 & 13.64 & 83.33 & 7.58 & 60.00 & \cellcolor{redbox}4.55 & \cellcolor{yellowbox}26.67 & -- & -- \\
Economic Law    & 66.28 & 72.97 & 78.57 & 86.49 & 76.19 & 83.78 & \cellcolor{greenbox}81.08 & 89.19 & 73.81 & 90.63 & 6.67  & 78.38 & 6.67  & \textbf{\cellcolor{bluebox}91.89} & \cellcolor{redbox}1.67 & 45.95 & 6.67 & \cellcolor{yellowbox}18.92 & -- & -- \\
Tax Law         & 67.11 & 73.81 & 75.76 & 83.33 & 69.23 & 76.19 & \cellcolor{greenbox}80.10 & \textbf{\cellcolor{bluebox}88.10} & 71.43 & 85.94 & 15.79 & 66.67 & 7.02  & 71.43 & \cellcolor{redbox}1.75 & 40.48 & 5.26 & \cellcolor{yellowbox}14.29 & -- & -- \\
Wealth Mgt.     & 45.45 & 50.00 & 58.06 & 63.89 & \cellcolor{greenbox}60.61 & 66.67 & 50.51 & 55.56 & 54.55 & \textbf{\cellcolor{bluebox}67.50} & \cellcolor{redbox}0.00  & 44.44 & \cellcolor{redbox}0.00  & 52.78 & \cellcolor{redbox}0.00 & 47.22 & \cellcolor{redbox}0.00 & \cellcolor{yellowbox}13.89 & -- & -- \\
\midrule
\textbf{All} & 80.81 & 68.45 & 73.42 & 84.80 & 69.37 & 80.46 & 81.06 & 80.00 & 76.32 & \textbf{92.00} & 11.18 & 69.14 & 7.57 & 73.14 & \textbf{2.96} & 49.71 & 4.61 & 19.43 & -- & -- \\
\bottomrule
\end{tabular}
}
\caption{\textbf{Accuracy (\%) across financial topics} for English CFA (top) and Chinese CPA (bottom). \textit{O} = original; \textit{R} = revised. For each row, best original is in \colorbox{green}{Best-O}, \colorbox{lightblue}{Best-R}, \colorbox{lightred}{Worst-O}, \colorbox{yellow}{Worst-R}; best per row is \textbf{bolded}.}
\label{tab:topic_cpa_reordered}

\end{table*}

\paragraph{Who Wins?} 
Unexpectedly, five general-purpose models overall perform the best, when financial models are the least (Tables~\ref{tab:type_cpa_reordered}, \ref{tab:topic_cpa_reordered}), with reasoning-enhanced models in the middle (\tabref{tab:reasoning_enhanced_models}).

The English CFA is led by Claude-Sonnet-3.5 under the missing-condition (89.62\%), followed by GPT-OSS-120B (89.10\%), GPT-5.1-mini (89.01\%), Qwen3-Max (88.83\%), DianJin-R1-32B (88.28\%), and Gemini-2.5-Flash (88.19\%) at the same level. 
In the Chinese CPA, Qwen3-Max establishes a significant lead by 92.00\%. Gemini-2.5-Flash (84.80\%) and GPT-OSS-120B (84.57\%) show comparable performance, as GPT-5.1-mini (80.81\%), Claude-Sonnet-3.5 (80.46\%) and DeepSeek-V3 (81.06\%) are similar.

\textbf{\textit{General Models:}} analyzing by question types in Table~\ref{tab:type_cpa_reordered}, GPT-5.1-mini sweeps all original categories in the CFA and DeepSeek-V3 in CPA. Revised tasks favor Claude-Sonnet-3.5 for CFA and Qwen3-Max for CPA.

Across topics in Table~\ref{tab:topic_cpa_reordered}, general models strikingly share identical high scores, with green/blue cells clustered instead of scattered. All general models gain 81.82\% for original and 90.00\% for revised in \textit{Economics}, 90.91\% ($O$) and 100.00\% ($R$) in \textit{Portfolio Management}, 79.66\% ($O$) in \textit{FSA}, and 71.43\% ($O$) in \textit{Fixed Income}. Such sameness may imply that \emph{(i)} general pre-training corpora cover these topics and \emph{(ii)} a common performance ceiling exists among state-of-the-art models.


\paragraph{What are Easy and What are Hard?}

Across question types for both O and R in Table~\ref{tab:type_cpa_reordered}, complex calculation is the hardest for all general models. Simple calculation and knowledge transfer \& application are comparably challenging. Notably, GPT-5.1-mini exhibits a systemic weakness in CPA, particularly in revised complex calculations. 

Regarding topic-specific performance in Table~\ref{tab:topic_cpa_reordered}, Fin-R1-7B achieves the highest scores in \textit{fixed income}, \textit{portfolio management}, and \textit{economic law}. In particular, XuanYuan3-70B, Fin-R1-7B, and CFGPT2-7B achieve a perfect precision of 100.0\% in \textit{portfolio management}, showing its easiness.  
Reasoning models in Table~\ref{tab:reasoning_full_breakdown} show similar trends.

\textbf{\textit{Financial Models:}} In the Chinese CPA, 4 times 0\% in \textit{knowledge application} and \textit{complex calculation} represent systemic ``dead zones'' for financial models. Under the wealth management topic, the finance-specific models also stand out by 4 times 0\%. These failures indicate that domain-specific supervised fine-tuning (SFT), though enhancing factual recall capabilities, has failed to cultivate the multi-step reasoning abilities that require for synthesising complex regulatory rules.

However, on the English CFA, XuanYuan3-70B, Fin-R1-7B, and CFGPT2-7B surprisingly attain their peak results in \textit{complex calculation} by 100\% and \textit{statistical methods} for revised variants. 

Overall, question difficulty is consistent across model architectures, which is determined by reasoning demands and information integration capabilities, rather than specific types or topics. Our correlation measurements also reflect the same finding: whether the questions is challenging is driven by information integration density ($r$ = 0.78) rather than surface features like question length ($\rho$ = 0.43) or numerical frequency ($\rho$ = 0.52). See measurement details in Appendix~\ref{app:spearman}, ~\ref{app:surface},~\ref{app:integration}.

\paragraph{Why Do Original and Revised Questions Behave Differently?}
The performance gap between original ($O$) and revised ($R$)  settings reveals how models respond to missing information. We examine accuracy changes across question types (Table \ref{tab:type_cpa_reordered}) and financial topics (Table \ref{tab:topic_cpa_reordered}). While the general models maintain a near-total monopoly on Best-$O$ results, a significant shift occurs in the revised setting, where approximately 1/5 of the Best-$R$ are captured by specialized financial models.

Among the general models, they tend to speculate rather than detecting missing information. Changing from original to revised, English CFA accuracies improve by 8\% to 12\%. Specifically, the accuracy of GPT-5.1-mini increased from 80.59\% to 89.01\% (+8.42\%), and DeepSeek-V3's accuracy rose from 73.55\% to 86.34\% (+12.79\%). However, the Chinese performance declines by 12\% (the accuracy of GPT-5.1-mini decreased from 80.81\% to 68.45\%), and the Worst-$R$ is 66.45\% with GPT-5.1-mini in the Chinese CPA \textit{complex calculation}. 

A natural question rises: \textbf{\textit{why these models can recognize the lack of information in English but not in Chinese?}} This difference is likely driven by how questions are expressed in the two languages. In Chinese, many words are highly context-dependent and can take on different meanings depending on surrounding cues. For example, words like \cn{走} (\textit{to walk}) can mean \textit{walk} or \textit{leave} in different contexts. Therefore, when a key condition is removed, the remaining words may still form a seemingly coherent sentence, but their meaning becomes ambiguous. Models are then easily misled by these remaining cues and tend to interpret the question in a plausible but unsupported way. In English, although context also matters, missing information more often triggers a different response. This can sometimes lead to higher accuracy, not because the model correctly recognizes the lack of information, but because the guessed assumption happens to align with the reference answer.

%
Table~\ref{tab:bilingual_example} presents paired bilingual examples that illustrate how the same model (GPT-5.1-mini) exhibits distinct failure modes in two languages.

The evaluation results on reasoning models further shows that the dead-zone pattern on Chinese CPA original questions is architecture-agnostic. GPT-OSS-120B (10.53\%), DianJin-R1-32B (13.82\%), and GPT-OSS-20B (8.88\%) all exhibit the same systematic failure, implying that the bottleneck is conceptual rather than computational.

By contrast, reasoning-focused financial models such as Fin-R1-7B show much larger gains under missing conditions (e.g.,\ +65.57\% on Chinese CPA), while earlier financial models such as DISC-FinLLM-13B and FinGPT-7B remain stagnant or decline. 

Specifically, zooming into English revised settings across question types in Table~\ref{tab:type_cpa_reordered}, XuanYuan3-70B, Fin-R1-7B, and CFGPT2-7B achieve 100.0\% accuracy in both \textit{complex calculation} and \textit{statistical methods}. These models similarly reach perfect scores in revised \textit{portfolio management} in Table \ref{tab:topic_cpa_reordered}, with Fin-R1-7B securing five Best-$R$ rankings across diverse topics. The jump from near-zero scores in the original \textit{complex calculations} to perfect performance in the revised setting shows that the issue is not math ability, but unclear concepts.

To further explore the two points above, we introduce an additional evaluation setting: None of the Above, testing whether models can genuinely identify which specific condition is missing.

\begin{table*}[t!]
\centering
\small
\setlength{\tabcolsep}{3.5pt}
\adjustbox{max width=\textwidth}{
\begin{tabular}{lccc|lccc|lccc}
\toprule
\multicolumn{4}{c|}{\textbf{General Models}} 
& \multicolumn{4}{c|}{\textbf{Financial Models}} 
& \multicolumn{4}{c}{\textbf{Reasoning-enhanced Models}} \\
\cmidrule(lr){1-4} \cmidrule(lr){5-8} \cmidrule(lr){9-12}
\textbf{Model} & \textbf{Chi} & \textbf{Eng} & \textbf{Avg} 
& \textbf{Model} & \textbf{Chi} & \textbf{Eng} & \textbf{Avg} 
& \textbf{Model} & \textbf{Chi} & \textbf{Eng} & \textbf{Avg} \\
\midrule
Gemini 2.5 Flash  & \cellcolor{red!20}48.78 & \cellcolor{red!20}72.38 & \cellcolor{red!20}60.58 
& Fin-R1-7B        & \cellcolor{green!20}77.10 & \cellcolor{green!20}84.50 & \cellcolor{green!20}81.10 
& DeepSeek-R1      & \cellcolor{red!20}73.31    & \cellcolor{red!20}66.93    & \cellcolor{red!20}70.12 \\
Claude 3.5 Sonnet & \cellcolor{red!20}37.36 & \cellcolor{red!20}70.11 & \cellcolor{red!20}53.73 
& XuanYuan3-70B    & \cellcolor{red!20}61.70   & \cellcolor{red!20}84.70   & \cellcolor{red!20}73.20  
& GPT-5.2-series   & \cellcolor{red!20}71.16    & \cellcolor{red!20}73.21    & \cellcolor{red!20}72.19 \\
Qwen3-Max         & \cellcolor{red!20}44.85 & \cellcolor{red!20}59.94 & \cellcolor{red!20}52.39 
& CFGPT2-7B        & \cellcolor{green!20}54.90 & \cellcolor{green!20}75.20 & \cellcolor{green!20}65.05 
& GPT-OSS-20B      & \cellcolor{yellow!25}67.43 & \cellcolor{yellow!25}86.65 & \cellcolor{yellow!25}77.04 \\
GPT-5.1-mini      & \cellcolor{red!20}30.64 & \cellcolor{red!20}68.23 & \cellcolor{red!20}49.44 
& DISC-FinLLM-13B  & \cellcolor{green!20}20.00 & \cellcolor{red!20}12.26   & \cellcolor{red!20}16.13  
& GPT-OSS-120B     & \cellcolor{yellow!25}84.57 & \cellcolor{yellow!25}89.10 & \cellcolor{yellow!25}86.84 \\
DeepSeek V3       & \cellcolor{red!20}36.21 & \cellcolor{red!20}56.87 & \cellcolor{red!20}46.54 
& FinGPT-7B        & \cellcolor{gray!20}--     & \cellcolor{red!20}3.50    & \cellcolor{red!20}3.50   
& DianJin-R1-32B   & \cellcolor{yellow!25}77.71 & \cellcolor{yellow!25}88.28 & \cellcolor{yellow!25}83.00 \\
\midrule
\textbf{Avg.} & \cellcolor{red!20}\textit{39.57} & \cellcolor{red!20}\textit{65.51} & \cellcolor{red!20}\textit{52.54} 
& \textbf{Avg.} & \cellcolor{green!20}\textit{53.43} & \cellcolor{red!20}\textit{52.03} & \cellcolor{red!20}\textit{47.80} 
& \textbf{Avg.} & \cellcolor{red!20}\textit{74.84} & \cellcolor{red!20}\textit{80.83} & \cellcolor{red!20}\textit{77.84} \\
\bottomrule
\end{tabular}
}
\caption{Cross-comparison of model accuracy (\%) on None-of-the-Above questions versus Revised accuracy, across three groups of models. Color coding: \colorbox{green!20}{Green:}NOTA $>$ Revised; \colorbox{red!20}{Red:}NOTA $<$ Revised; \colorbox{yellow!25}{Yellow:}NOTA $=$ Revised (model treats NOTA options identically to standard distractors).}
\label{tab:combined_nota_comparison}
\end{table*}

\paragraph{Why Do LLMs Make Errors?}
The additional evaluation reveals that many LLM errors stem from an inability to reason about which specific condition is missing.

\textbf{\textit{General-purpose models}} such as GPT-5.1-mini and DeepSeek-V3 often respond with high confidence even when key information is missing.
When models are asked not only to choose an answer but to identify what information is absent, general models often fail to give a concrete explanation. Instead, they fall back on vague assumptions or restate parts of the question. This behavior indicates that the error is not simply a wrong final choice, but defaults to guessing based on learned patterns.



This behavior is confirmed quantitatively by the substantial accuracy drop from the Revised to the None of the Above (NOTA) setting shown in Table~\ref{tab:combined_nota_comparison}. If models were reasoning strictly from the given information, performance should remain stable when the correct option is replaced by \textit{None of the Above}. However, drops occur across general models, indicating that their correct answers in the Revised setting often come from guessing by filling in missing information, rather than reasoning from what is actually given.

In other words, models prioritize producing a plausible completion over verifying whether the question contains sufficient information. When a key condition is missing, they substitute it with a common scenario learned from training data. This makes the answer appear coherent, even though it is no longer supported by the problem statement.

\textbf{\textit{Financial models}} such as DISC-FinLLM-13B and FinGPT-7B perform extremely poorly, with accuracy below 20.00\%.
This outcome is counter-intuitive, as these models are explicitly fine-tuned for financial tasks and possess substantial domain knowledge.
A closer inspection suggests that this poor performance is not due to a lack of financial knowledge.
Instead, domain-specific terminology in the questions, such as "Processing Agreement" or "Consumption Tax," appears to trigger aggressive knowledge retrieval.
Once this retrieval process is activated, the models tend to commit to forced calculations or rule-based procedures, even when the information provided is insufficient to support any option. As a result, these models are more likely to select an answer rather than recognize that no option can be justified, which leads to particularly low accuracy in all settings.

In other words, early financial models such as DISC-FinLLM-13B and FinGPT-7B are trained on financial data, they mainly learn to recognize financial terms, and this helps them reduce errors caused by unfamiliar terminology and better identify the meaning of a question.
However, these models are not trained to recognize when the information provided is insufficient, hence, still lack the ability to detect when a question itself cannot be answered.

Reasoning-focused models such as Fin-R1-7B also face a new challenge. 
Although Fin-R1-7B has learned to reject answers when information is insufficient, it sometimes struggles to decide where to stop.

We observe that Fin-R1-7B frequently identifies the missing condition correctly in its reasoning trace but continues generating for \textbf{\textit{200--300 additional tokens}} exploring hypothetical scenarios not implied by the question. 
Its reasoning trajectory shows that after the model correctly recognizes the core principles, it still continues reasoning, such as considering unlikely financial cases that are not implied by the question.

\textbf{\textit{Take-Away:}} 
The ability to say \textit{No} is essential for handling missing conditions, but without a clear stopping criterion, it can also trigger unnecessary complexity.
This also mirrors a common challenge in human decision-making, where caution and thoroughness sometimes lead to overthinking.

\subsection{How Do Models Behave When No Answer is Justified}
Based on the revised subset, we replace the correct option with ``none of the above'' as the final choice, to assess whether models can truly identify missing information when no answer is justified.

\paragraph{Accuracy Decreases for Most Models When Models Are Forced to Choose NOTA}
More apparent declines happen among general-purpose models. Financial-specific models witness majority of rises in Chinese questions in Table~\ref{tab:combined_nota_comparison}.\footnote{The financial models and the reasoning-enhanced models in Table~\ref{tab:combined_nota_comparison} were evaluated on different GPU hardware, which may account for minor numerical variances in the accuracies.}

\paragraph{Most Models Still Try to Answer Even When No Answer Is Justified}

Both finance-specific models and general frontier models struggle when \emph{None of the above} is the correct answer.
In Table~\ref{tab:combined_nota_comparison}, models such as GPT-5.1-mini and DeepSeek V3 show large accuracy drops, especially on Chinese questions.
These errors do not stem from a lack of reasoning ability, but from a tendency to keep selecting an answer even when the question cannot be resolved with the given information.

Empirically, general-purpose models tend to over-commit by selecting an answer even when the question lacks a clear logical basis.
This behavior likely comes from training objectives that reward producing a response rather than recognizing when a problem cannot be solved.
As a result, these models fail not because they cannot reason, but because they do not know when to stop reasoning.

However, the strong performance of Fin-R1-7B shows that this limitation is not shared by all models.
Unlike other models, Fin-R1-7B does not treat the \emph{None of the above} option as a fallback choice.
Instead, it consistently checks whether the information provided in the question is sufficient to justify any of the listed options. Its reasoning outputs indicate that the model actively looks for missing inputs before committing to an answer.
For example, in tax-related questions, Fin-R1-7B explicitly notes the absence of required quantities, such as a comparable market price, and refrains from selecting an option when such information is missing.
As Fin-R1-7B is designed to prioritize condition checking before answer selection, which directly counteracts the tendency to over-answer encouraged by RLHF.
This design also prevents the retrieval-heavy calculation mode triggered by financial terminology from turning into forced commitments, so the model can choose NOTA when the question is underspecified.

\paragraph{Why Accuracy Overestimates Reliability in NOTA}

In NOTA settings, selecting the correct option alone is insufficient to indicate reliable decision-making.
When no answer is justified, accuracy can be artificially inflated by models that avoid making a clear choice without understanding why the problem cannot be resolved.

\begin{table}[t!]
\centering
\small
\adjustbox{max width=\columnwidth}{
\setlength{\tabcolsep}{3pt} 
\begin{tabular}{l rrr rrr}
\toprule
\multirow{2}{*}{\textbf{Model}} & \multicolumn{3}{c}{\textbf{English CFA} (367)} & \multicolumn{3}{c}{\textbf{Chinese CPA} (175)} \\
\cmidrule(lr){2-4} \cmidrule(lr){5-7}
& \textit{Ans.} & \textit{Reas.} & \textit{Conf.} & \textit{Ans.} & \textit{Reas.} & \textit{Conf.} \\
\midrule
XuanYuan3-70B & 84.7 & 8.4 & 92.9 & 61.7 & 5.1 & 87.9 \\
Fin-R1-7B & 84.5 & 42.5 & 93.8 & 77.7 & 41.1 & 90.3 \\
CFGPT2-7B & 75.2 & 8.5 & 82.6 & 54.9 & 11.4 & 81.0 \\
DISC-13B & 12.3 & 2.3 & 24.2 & 20.0 & 18.9 & 65.2 \\
FinGPT-7B & 3.5 & 0.0 & 50.3 & -- & -- & -- \\
\midrule
GPT-OSS-20B & 86.6 & 12.3 & 90.7 & 67.4 &  4.0 & 79.2 \\
GPT-OSS-120B & 89.1 & 20.7 & 95.7 & 84.6 &  1.1 & 91.5 \\
DianJin-R1-32B & 88.3 & 29.2 & 93.7 & 77.7 &  5.7 & 77.8 \\
\bottomrule
\end{tabular}
}
\caption{Open-source models' NOTA results: Answer Acc vs.\ Reasoning Acc gap, along with confidence.}
\label{tab:answer-vs-reasoning-acc-gap}
\end{table}
To address this, we examine whether a model’s reasoning supports the same conclusion as its final answer. To distinguish principled abstentions from lucky guesses, we evaluate every NOTA response against two complementary metrics: \textit{Answer Accuracy} (whether NOTA is selected) and \textit{Reasoning Accuracy} (whether the explanation correctly identifies the specific missing premise).

In \tabref{tab:answer-vs-reasoning-acc-gap}, eight LLMs consistently exhibit a significant gap between answer accuracy and reasoning accuracy. The smallest gap occurs in reasoning-focued financial model Fin-R1-7B. It maintains relatively higher consistency between answers and explanations, identifying which required condition is missing before selecting NOTA.

By contrast, some financial models and reasoning models appear to perform reasonably well based on answer accuracy alone, but their explanations fail to support the chosen answer.
For example, XuanYuan3-70B, GPT-OSS-120B/20B and DianJin-R1 drop from >85\% answer accuracy to 8-30\% for reasoning on English NOTA questions. In these cases, the model selects NOTA simply because it is a safe option, not because it understands that the question cannot be answered.

We further observe a degenerate response pattern in some models, where uncertainty is expressed by selecting multiple options at once (\textit{e.g.,\ AB or ABCD}).
Although this behavior may seem cautious, it does not represent a valid financial decision because it offers no clear conclusion.
Taken together, financial applications require models to reject an answer for clear and traceable reasons, rather than by chance.

\section{Conclusions}

In this paper, we show that evaluating financial LLMs solely by answer accuracy misses a key aspect of reliability: knowing when not to answer. 
Models often answer despite missing information, especially in Chinese, where terminology hides missing conditions and triggers more assumptions.
Financial models reduce blind guessing but still struggle to identify what is missing, while reasoning-focused models improve condition detection but may over-reason. 
Scaling alone does not solve this: larger models achieve high answer accuracy but low reasoning accuracy, increasing over-commitment without improving condition checking.
We therefore argue that reliable financial LLMs must satisfy three criteria: answer correctly when information is sufficient, identify missing conditions when it is not, and abstain when no valid answer exists. 
Only models that perform consistently across these settings in both languages are dependable in practice.

\section*{Limitations and Future Work}

A main limitation of this work is that we do not modify or train any models. Our study can show where failures come from, but we do not directly demonstrate how much fine-tuning can fix them.  Because our evaluation uses multiple-choice questions, the same behavior may show up in different ways when models respond free-form questions or interact with users.

These limits point to a clear next step. As our results suggests that models either guess, or they pick a safe option without being able to explain what is missing, or they reason too far.  A practical solution is a unified training pipeline that separates what should be learned from what should be controlled. First, specialized SFT should teach professional knowledge and stable reasoning formats, so the model learns what variables and rules matter in common financial tasks. Then logic-driven RL should train the model’s behavior when conditions are missing. At this stage, the model should be rewarded for explicitly pointing out what information is missing, and penalized for guessing, making up assumptions, or reasoning too far beyond what the question supports.

We also plan to move beyond passive rejection. When a question lacks a key variable, the model should not only say ''\textit{This question has no answer}'', but also state what information it needs and ask for it. This would build a interactive workflow for users. We will also broaden the dataset beyond accounting and tax law to cover risk management, securities regulation, and quantitative finance, so that the same evaluation logic can test whether models stay reliable across the full range of real financial work.

\paragraph{Ethical Statement}

This work is intended to improve the safe and responsible use of large language models in financial applications. All datasets used in this study are constructed from publicly accessible materials or manually rewritten questions and do not include proprietary data, confidential information, or verbatim content from protected examinations. Our evaluation highlights a specific risk relevant to high-stakes domains: confident model outputs in situations where information is insufficient to justify a determinate answer. By explicitly identifying and measuring this behaviour, our work aims to reduce the likelihood of uncritical deployment of LLMs in financial decision-making, rather than to encourage automation of such decisions.

We emphasise that the proposed benchmark and any derivative artefacts, such as distilled small-parameter models, are intended for research and risk-assessment purposes only. Any deployment of language models in real financial, regulatory, or advisory contexts must prioritize the auditability of reasoning chains, involve appropriate human oversight, and comply with applicable legal and ethical standards. We aim to support the development of "logic-first" assistance tools rather than fully autonomous systems for high-stakes financial judgments.


\bibliographystyle{acl_natbib}
\bibliography{main}

\clearpage
\appendix

\begin{table*}[t!]
\centering
\small
\adjustbox{max width=\textwidth}{
\setlength{\tabcolsep}{3pt}
\begin{tabular}{@{}l*{5}{cc}@{}}
\toprule
& \multicolumn{10}{c}{\textbf{Reasoning-enhanced Models}} \\
\cmidrule(lr){2-11}
& \multicolumn{2}{c}{DeepSeek-R1} & \multicolumn{2}{c}{GPT-5.2} & \multicolumn{2}{c}{GPT-OSS-20B} & \multicolumn{2}{c}{GPT-OSS-120B} & \multicolumn{2}{c}{DianJin-32B} \\
\cmidrule(lr){2-3}\cmidrule(lr){4-5}\cmidrule(lr){6-7}\cmidrule(lr){8-9}\cmidrule(lr){10-11}
\textbf{Category} & \textit{O} & \textit{R} & \textit{O} & \textit{R} & \textit{O} & \textit{R} & \textit{O} & \textit{R} & \textit{O} & \textit{R} \\
\midrule
\multicolumn{11}{c}{\texttt{English CFA}} \\
\midrule
\multicolumn{11}{l}{\textit{By Question Type}} \\
\cmidrule(l){1-1}
Conceptual     & \cellcolor{redbox}79.13 & \cellcolor{yellowbox}71.34 & 81.45 & 74.42 & 83.87 & 83.72 & \cellcolor{greenbox}85.48 & \textbf{\cellcolor{bluebox}86.05} & 80.65 & \textbf{\cellcolor{bluebox}86.05} \\
Simple Calc.   & 71.67 & \cellcolor{yellowbox}68.53 & 75.71 & 71.26 & \cellcolor{redbox}63.33 & 84.92 & \textbf{\cellcolor{greenbox}93.33} & \cellcolor{bluebox}87.71 & 66.67 & 87.15 \\
Complex Calc.  & 73.49 & \cellcolor{yellowbox}76.56 & 78.13 & 81.25 & \cellcolor{redbox}60.00 & \textbf{\cellcolor{bluebox}100.00} & \cellcolor{greenbox}86.67 & \textbf{\cellcolor{bluebox}100.00} & \cellcolor{redbox}60.00 & \textbf{\cellcolor{bluebox}100.00} \\
Comp. Judg.    & 76.65 & \cellcolor{yellowbox}70.47 & 80.12 & 74.83 & \cellcolor{redbox}68.32 & 88.37 & 78.22 & \textbf{\cellcolor{bluebox}89.53} & \cellcolor{greenbox}84.16 & \textbf{\cellcolor{bluebox}89.53} \\
Knowledge App. & 72.09 & \cellcolor{yellowbox}67.38 & \cellcolor{greenbox}76.04 & 71.59 & 48.84 & 87.50 & 56.98 & \textbf{\cellcolor{bluebox}100.00} & \cellcolor{redbox}44.19 & 87.50 \\
Stats Methods  & \cellcolor{redbox}80.00 & \cellcolor{yellowbox}78.43 & 84.29 & 82.13 & 88.57 & 91.49 & \cellcolor{greenbox}90.00 & \textbf{\cellcolor{bluebox}93.62} & \cellcolor{greenbox}90.00 & 91.49 \\
\cmidrule(l){1-1}
\multicolumn{11}{l}{\textit{By Topic}} \\
\cmidrule(l){1-1}
Corp. Finance  & \cellcolor{redbox}77.50 & \cellcolor{yellowbox}71.76 & \cellcolor{greenbox}81.25 & 75.34 & 80.00 & \textbf{\cellcolor{bluebox}90.00} & 80.00 & \textbf{\cellcolor{bluebox}90.00} & 80.00 & \textbf{\cellcolor{bluebox}90.00} \\
Derivatives    & 68.89 & \cellcolor{yellowbox}63.27 & \cellcolor{greenbox}73.63 & 67.35 & \cellcolor{redbox}60.00 & 88.00 & 71.11 & 88.00 & 62.22 & \textbf{\cellcolor{bluebox}96.00} \\
Economics      & 72.22 & \cellcolor{yellowbox}67.69 & \cellcolor{greenbox}76.87 & 70.38 & 66.67 & \textbf{\cellcolor{bluebox}90.00} & 66.67 & \textbf{\cellcolor{bluebox}90.00} & \cellcolor{redbox}60.00 & \textbf{\cellcolor{bluebox}90.00} \\
Equity Inv.    & 77.97 & \cellcolor{yellowbox}72.43 & 81.33 & 75.96 & \cellcolor{redbox}74.58 & \textbf{\cellcolor{bluebox}88.46} & \cellcolor{greenbox}83.05 & \textbf{\cellcolor{bluebox}88.46} & 76.27 & \textbf{\cellcolor{bluebox}88.46} \\
FSA            & \cellcolor{greenbox}76.74 & \cellcolor{yellowbox}71.48 & 80.24 & 74.81 & \cellcolor{redbox}74.42 & 85.19 & \cellcolor{redbox}74.42 & \textbf{\cellcolor{bluebox}88.89} & \cellcolor{greenbox}76.74 & \textbf{\cellcolor{bluebox}88.89} \\
Fixed Income   & 63.27 & \cellcolor{yellowbox}58.16 & \cellcolor{greenbox}67.35 & 61.22 & 53.06 & \textbf{\cellcolor{bluebox}78.57} & \cellcolor{greenbox}67.35 & \textbf{\cellcolor{bluebox}78.57} & \cellcolor{redbox}46.94 & \textbf{\cellcolor{bluebox}78.57} \\
Other Inv.     & 75.53 & \cellcolor{yellowbox}70.37 & 78.59 & 73.46 & \cellcolor{redbox}68.09 & 88.89 & \cellcolor{greenbox}81.91 & \textbf{\cellcolor{bluebox}90.12} & 79.79 & 88.89 \\
Portfolio Mgt  & \cellcolor{redbox}90.91 & \cellcolor{yellowbox}88.24 & 95.45 & 92.71 & \textbf{\cellcolor{greenbox}100.00} & \textbf{\cellcolor{bluebox}100.00} & \textbf{\cellcolor{greenbox}100.00} & \textbf{\cellcolor{bluebox}100.00} & \textbf{\cellcolor{greenbox}100.00} & \textbf{\cellcolor{bluebox}100.00} \\
Quant Analysis & \cellcolor{redbox}78.72 & \cellcolor{yellowbox}73.47 & 82.98 & 76.83 & 85.11 & 88.89 & \cellcolor{greenbox}89.36 & \textbf{\cellcolor{bluebox}91.11} & \cellcolor{greenbox}89.36 & 88.89 \\
\midrule
\textbf{All}   & 76.50 & 72.81 & 83.80 & 79.23 & 69.67 & 86.65 & 78.42 & \textbf{89.10} & 72.95 & 88.28 \\
\midrule
\multicolumn{11}{c}{\texttt{Chinese CPA}} \\
\midrule
\multicolumn{11}{l}{\textit{By Question Type}} \\
\cmidrule(l){1-1}
Conceptual     & 80.34 & \cellcolor{yellowbox}76.82 & \cellcolor{greenbox}82.18 & 78.64 & 10.08 & 90.00 & \cellcolor{redbox}8.40 & \textbf{\cellcolor{bluebox}100.00} & 15.13 & 90.00 \\
Simple Calc.   & 78.57 & 74.29 & \textbf{\cellcolor{greenbox}80.14} & 75.96 & \cellcolor{redbox}2.44 & \cellcolor{yellowbox}49.06 & 4.88 & \cellcolor{bluebox}69.81 & 4.88 & 62.26 \\
Complex Calc.  & 76.92 & \cellcolor{bluebox}72.73 & \textbf{\cellcolor{greenbox}78.85} & 75.19 & \cellcolor{redbox}0.00 & -- & \cellcolor{redbox}0.00 & -- & \cellcolor{redbox}0.00 & -- \\
Comp. Judg.    & 78.16 & \cellcolor{yellowbox}74.58 & \cellcolor{greenbox}80.23 & 76.34 & \cellcolor{redbox}4.60 & 77.78 & 9.20 & \textbf{\cellcolor{bluebox}91.92} & 13.79 & 84.85 \\
Knowledge App. & 76.34 & 71.21 & \cellcolor{greenbox}78.57 & 73.28 & 14.29 & \cellcolor{yellowbox}33.33 & 28.57 & 66.67 & \cellcolor{redbox}0.00 & \textbf{\cellcolor{bluebox}100.00} \\
Stats Methods  & 79.13 & 75.45 & \textbf{\cellcolor{greenbox}81.25} & 77.91 & \cellcolor{redbox}18.75 & \cellcolor{yellowbox}50.00 & 20.83 & \cellcolor{bluebox}80.00 & 20.83 & 70.00 \\
\cmidrule(l){1-1}
\multicolumn{11}{l}{\textit{By Topic}} \\
\cmidrule(l){1-1}
Accounting     & 78.41 & 73.67 & \textbf{\cellcolor{greenbox}80.23} & 75.42 & 6.74 & \cellcolor{yellowbox}36.67 & \cellcolor{redbox}5.62 & \cellcolor{bluebox}66.67 & 7.87 & 60.00 \\
Auditing       & 80.76 & \cellcolor{yellowbox}91.52 & \cellcolor{greenbox}83.94 & 93.18 & \cellcolor{redbox}16.67 & \textbf{\cellcolor{bluebox}93.33} & 25.76 & \textbf{\cellcolor{bluebox}93.33} & 36.36 & \textbf{\cellcolor{bluebox}93.33} \\
Economic Law   & 79.39 & 81.43 & \cellcolor{greenbox}81.67 & 83.47 & \cellcolor{redbox}3.33 & \cellcolor{yellowbox}70.27 & 6.67 & \textbf{\cellcolor{bluebox}91.89} & 8.33 & 89.19 \\
Tax Law        & 79.82 & \cellcolor{yellowbox}75.64 & \cellcolor{greenbox}82.46 & 77.14 & \cellcolor{redbox}7.02 & 83.33 & \cellcolor{redbox}7.02 & \textbf{\cellcolor{bluebox}92.86} & 8.77 & 88.10 \\
Wealth Mgt.    & 77.65 & 73.16 & \textbf{\cellcolor{greenbox}78.29} & 74.38 & 12.50 & \cellcolor{yellowbox}50.00 & 6.25 & \cellcolor{bluebox}75.00 & \cellcolor{redbox}3.12 & 55.56 \\
\midrule
\textbf{All}   & 79.27 & 77.54 & 80.24 & 75.39 & 8.88 & 67.43 & 10.53 & \textbf{84.57} & 13.82 & 77.71 \\
\bottomrule
\end{tabular}
}
\caption{\textbf{Accuracy (\%) of reasoning-enhanced models across question types and financial topics} for English CFA (top) and Chinese CPA (bottom). \textit{O} = original; \textit{R} = revised. For each row, best original is in \colorbox{greenbox}{Best-O}, best revised is in \colorbox{bluebox}{Best-R}, worst original is in \colorbox{redbox}{Worst-O}, and worst revised is in \colorbox{yellowbox}{Worst-R}; best per row is \textbf{bolded}. Ties are all highlighted.}
\label{tab:reasoning_full_breakdown}
\end{table*}

\begin{table*}[t]
\centering
\small
\begin{tabular}{p{0.08\textwidth}p{0.38\textwidth}p{0.38\textwidth}}
\toprule
& \textbf{English Version} & \textbf{Chinese Version} \\
\midrule
\textbf{Question} 
& When evaluating audit evidence, if a CPA discovers a 
material misstatement caused by fraud, which of the 
following measures should be taken? 
A. Issue a modified audit report. 
B. Re-evaluate the risk of material misstatement due to fraud. 
C. Reconsider the reliability of previously obtained audit evidence. 
D. Expand the scope of audit procedures.
& \cn{在评价审计证据时，注册会计师发现由于舞弊导致的重大错报，
应当采取的措施包括（）。
A. 出具非无保留意见的审计报告
B. 重新评估舞弊导致的重大错报风险
C. 重新考虑此前获取的审计证据的可靠性
D. 扩大审计程序的范围} \\
\midrule
\textbf{Correct answer} 
& BC & BC \\
\midrule
\textbf{Missing condition} 
& The applicable auditing standard (e.g., ISA 240 vs.\ 
PCAOB AS 2401) is unspecified, leaving the scope of 
required responses underdetermined. 
& \cn{未指明适用的审计准则体系（如中国注册会计师审计准则第
1141号），导致所要求的应对措施范围无法唯一确定。}\\
\midrule
\textbf{Model output (GPT-5.1-mini)} 
& \textit{Reason:} ``When fraud is detected, the auditor 
must re-evaluate the fraud risk assessment and 
reconsider the reliability of prior evidence. 
Issuing a modified report (A) is premature at this 
stage, and expanding audit scope (D) is not 
automatically required.''
\newline\textit{Answer:} BC \quad \textit{Conf:} 85
& \textit{Reason:} \cn{注册会计师发现舞弊导致的重大错报时，
应当重新评估相关风险、重新考虑审计证据可靠性、扩大审计
程序范围，并在必要时出具非无保留意见的审计报告。}
\newline\textit{Answer:} ABCD \quad \textit{Conf:} 91 \\
\midrule
\textbf{Failure mode} 
& \cellcolor{yellow!15}Assumption filling: model silently 
assumes ISA 240 scope and excludes A and D as 
conditionally required, arriving at BC by coincidence 
rather than by recognizing the missing standard.
& \cellcolor{red!20}Aggressive retrieval: terminology 
(\cn{``舞弊''、``重大错报''}) triggers exhaustive rule 
recall, causing the model to include all options 
regardless of conditionality, resulting in ABCD. \\
\bottomrule
\end{tabular}
\caption{Paired bilingual example illustrating cross-lingual 
failure modes of GPT-5.1-mini on the same audit question.
\colorbox{yellow!15}{English}: model arrives at correct answer 
BC by silently assuming a default auditing standard, not by 
detecting the missing condition.
\colorbox{red!20}{Chinese}: professional terminology triggers 
exhaustive knowledge retrieval, producing the overcomplete 
answer ABCD.}
\label{tab:bilingual_example}
\end{table*}

\clearpage
\section{Dataset Curation}
\subsection{Formal Definition of Condition-Missing Construction}

We formalize the construction of condition-missing questions to 
clarify how missing information is introduced in a controlled manner.

\paragraph{Full-condition formulation.}
Each full-condition financial reasoning question can be abstracted 
as a logical implication
\begin{equation}
(P_1 \land P_2 \land \dots \land P_n) \Rightarrow C,
\end{equation}
where $\{P_i\}$ denotes the set of premises explicitly stated in 
the question, such as numerical assumptions, regulatory standards, 
time horizons, or accounting rules, and $C$ denotes the unique 
justified conclusion. Under full conditions, the premise set is 
sufficient to determine $C$ without ambiguity.

\paragraph{Condition removal.}
To construct a condition-missing question, we remove one or more 
logically necessary premises from the original set. Let 
$\mathcal{P}^\ast \subset \{P_1,\dots,P_n\}$ denote the subset 
of premises identified as necessary for determining the conclusion. 
The revised question is defined by the reduced premise set
\begin{equation}
\{P_i\} \setminus \mathcal{P}^\ast.
\end{equation}

\paragraph{Coherence constraint.}
After removal, the remaining premises must still admit at least 
one coherent interpretation. Formally, we require
\begin{equation}
\exists \, \mathcal{M} \quad \text{s.t.} \quad 
\mathcal{M} \models \bigwedge_{P_i \in \{P\} \setminus \mathcal{P}^\ast} P_i,
\end{equation}
ensuring that the revised question remains grammatically and 
semantically well-formed, rather than logically inconsistent.

\paragraph{Non-uniqueness constraint.}
At the same time, the remaining premises must no longer uniquely 
determine the conclusion. Specifically, we require the existence 
of multiple internally consistent interpretations
\begin{equation}
\mathcal{M}_j \models \bigwedge_{P_i \in \{P\} \setminus \mathcal{P}^\ast} P_i
\quad \text{and} \quad
\mathcal{M}_j \models C_j,
\end{equation}
with
\begin{equation}
C_j \neq C_{j'},
\end{equation}
for some $j \neq j'$. This condition guarantees that no single 
answer can be justified given the available information.

\paragraph{Functional tagging.}
Each removed premise $P_k \in \mathcal{P}^\ast$ is annotated by 
its functional role
\begin{equation}
\tau(P_k) \in \{\textit{numerical}, \textit{regulatory}, 
\textit{temporal}, \textit{standard}\},
\end{equation}
to ensure that only logically necessary information is removed, 
rather than incidental details.

\paragraph{Implications for accuracy and model behavior.}
In original questions, accuracy directly reflects whether the 
model successfully derives the unique justified conclusion $C$. 
In revised questions, however, accuracy depends on whether the 
model's implicit assumptions happen to align with the reference 
answer retained from the original task. Formally, a correct 
prediction on a revised question satisfies
\begin{equation}
\hat{C} = C_{\text{ref}},
\end{equation}
even though
\begin{equation}
\{P_i\} \setminus \mathcal{P}^\ast \;\not\models\; C_{\text{ref}}.
\end{equation}
This explains why some models improve after condition removal 
while others degrade: performance changes reflect differences in 
fallback assumptions and inductive biases, rather than differences 
in reasoning ability. Consequently, accuracy ceases to be a 
reliable indicator of reasoning quality once questions are no 
longer well-posed. Together, these constraints ensure that 
condition-missing questions differ from full-condition questions 
not in computational difficulty, but in epistemic status: the 
task shifts from computing a correct answer to recognizing that 
no uniquely justified answer can be derived.

\subsection{Formalization of Abstention Under Missing Correct Answers}

This appendix provides a formal explanation of the experiment in 
Section~4, where the correct option in condition-missing questions 
is replaced by \emph{``None of the above''} (English) or 
\cn{``以上都不是''} (Chinese).

\paragraph{Problem setup.}
A full-condition financial reasoning question can be abstracted 
as a set of premises and a target conclusion,
\begin{equation}
(P_1 \land P_2 \land \dots \land P_n) \;\models\; C,
\end{equation}
where the premises jointly support a unique, justified conclusion. 
In the condition-missing setting, one or more logically necessary 
premises are removed. The resulting task is characterized by
\begin{equation}
(P_1 \land \dots \land P_{k-1} \land P_{k+1} \land \dots \land P_n) 
\;\not\models\; C,
\end{equation}
meaning that the remaining information is insufficient to justify 
any single concrete answer.

\paragraph{Implicit assumption completion.}
Despite this underdetermination, models often still select a 
concrete option. This behavior can be understood as implicitly 
completing the missing information with a default assumption. 
Formally, the model selects an internal completion $\mathcal{M}_j$ 
such that
\begin{equation}
\mathcal{M}_j \models \bigwedge_{i \neq k} P_i
\quad \text{and} \quad
\mathcal{M}_j \models C_j,
\end{equation}
where $C_j$ corresponds to one of the provided answer options. 
If $C_j$ happens to match the reference answer retained from the 
original full-condition question, the prediction is counted as 
correct, even though the conclusion is not logically supported 
by the observed premises alone.

\paragraph{Abstention as the rational strategy.}
When the correct option is explicitly set to \emph{``None of the 
above''}, this shortcut is no longer available. In this setting, 
a rational response requires recognizing that none of the concrete 
options is justified. Formally, the correct decision rule is
\begin{equation}
\forall j,\;\; \{P_i\} \not\models C_j
\quad \Rightarrow \quad
\text{select NOTA}.
\end{equation}
This formulation makes explicit that the task is no longer to 
identify the best answer, but to assess whether any answer can 
be supported at all. Empirically, however, models frequently 
violate this condition by selecting a specific option even when 
no option is entailed.

\paragraph{Interpretation.}
The sharp accuracy drop observed in the \emph{None-of-the-above} 
setting therefore indicates that many correct answers in the 
revised setting were achieved through implicit assumption 
alignment rather than principled reasoning. This experiment 
isolates a structural limitation of current LLMs: they lack a 
robust mechanism for recognizing when available information is 
insufficient and reasoning should be suspended rather than 
completed.

\begin{table*}[t]
\centering
\small
\renewcommand{\arraystretch}{1.2}
\setlength{\tabcolsep}{5pt}
\begin{tabular}{@{}p{0.08\linewidth}p{0.08\linewidth}p{0.79\linewidth}@{}}
\toprule
\textbf{Dataset} & \textbf{Type} & \textbf{Content} \\
\midrule

\multirow{6}{*}{\rotatebox[origin=c]{90}{\parbox{1.5cm}{\centering Original\\(Chinese)}}} 
& Topic & Auditing \\
& Q & \cn{下列关于管理层编制财务报告的说法中，正确的有（）。} \\
& Opts & \cn{A. 根据财务报告编制基础编制，不能运用自身判断 | B. 根据相关法律法规的规定确定适用的财务报告编制基础 | C. 根据适用的财务报告编制基础编制财务报表 | D. 在财务报表中对使用的财务报告编制基础作出恰当的说明} \\
& \cellcolor{yellow!20}\textit{Gold} & \cellcolor{yellow!20}\textbf{\textit{BCD}} \\
& \cellcolor{yellow!20}\textit{Rsn} & \cellcolor{yellow!20}\textbf{\textit{\cn{管理层在编制财务报表时应严格按照财务报告编制基础，但是也存在需要根据使用的财务报告编制基础运用判断做出合理会计估计的情况，不是完全不运用自身判断。}}} \\
\midrule

\multirow{6}{*}{\rotatebox[origin=c]{90}{\parbox{1.5cm}{\centering Missing\\(Chinese)}}} 
& Topic & Accounting \\
& Q & \cn{若要计算委托加工业务中受托方应代收代缴的消费税额，应补充下列哪项关键信息（）。} \\
& Opts & \cn{A. 该应税物资在受托方是否存在同类消费品的销售价格 | B. 该批原材料是否属于自产产品 | C. 加工费是否以现金方式结算 | D. 委托方是否具有一般纳税人资格} \\
& \cellcolor{yellow!20}\textit{Gold} & \cellcolor{yellow!20}\textbf{\textit{A}} \\
& \cellcolor{yellow!20}\textit{Rsn} & \cellcolor{yellow!20}\textbf{\textit{\cn{需要知道是否有同类消费品销售价格来确定计税依据。}}} \\
\midrule

\multirow{6}{*}{\rotatebox[origin=c]{90}{\parbox{1.5cm}{\centering Original\\(English)}}} 
& Topic & Financial Statement Analysis \\
& Q & Which of the following is most likely to have been included in Sea Ltd's SEC registration statement? \\
& Opts & A. Underwriters' fairness opinion | B. Assessment of risk factors | C. Projected cash flows and earnings \\
& \cellcolor{yellow!20}\textit{Gold} & \cellcolor{yellow!20}\textbf{\textit{B}} \\
& \cellcolor{yellow!20}\textit{Rsn} & \cellcolor{yellow!20}\textbf{\textit{SEC registration statements must include risk factor assessments.}} \\
\midrule

\multirow{6}{*}{\rotatebox[origin=c]{90}{\parbox{1.5cm}{\centering Missing\\(English)}}} 
& Topic & Derivatives Investment \\
& Q & For partnerships to achieve pass-through taxation, which structural characteristic must be present? \\
& Opts & A. Entity level taxation | B. Individual partner level only | C. Both entity and individual | D. No taxation \\
& \cellcolor{yellow!20}\textit{Gold} & \cellcolor{yellow!20}\textbf{\textit{B}} \\
& \cellcolor{yellow!20}\textit{Rsn} & \cellcolor{yellow!20}\textbf{\textit{Pass-through taxation means income is taxed only at individual partner level.}} \\

\bottomrule
\end{tabular}
\caption{Question examples from Original and Missing datasets. Gold answers and reasons are highlighted in yellow and displayed in \textbf{\textit{bold italic}}. ``Q'' = Question, ``Opts'' = Options, ``Gold'' = Gold Answer, ``Rsn'' = Reason.}
\label{tab:question_examples}
\end{table*}

\begin{figure*}[!ht]
  \centering
  \includegraphics[width=0.95\linewidth]{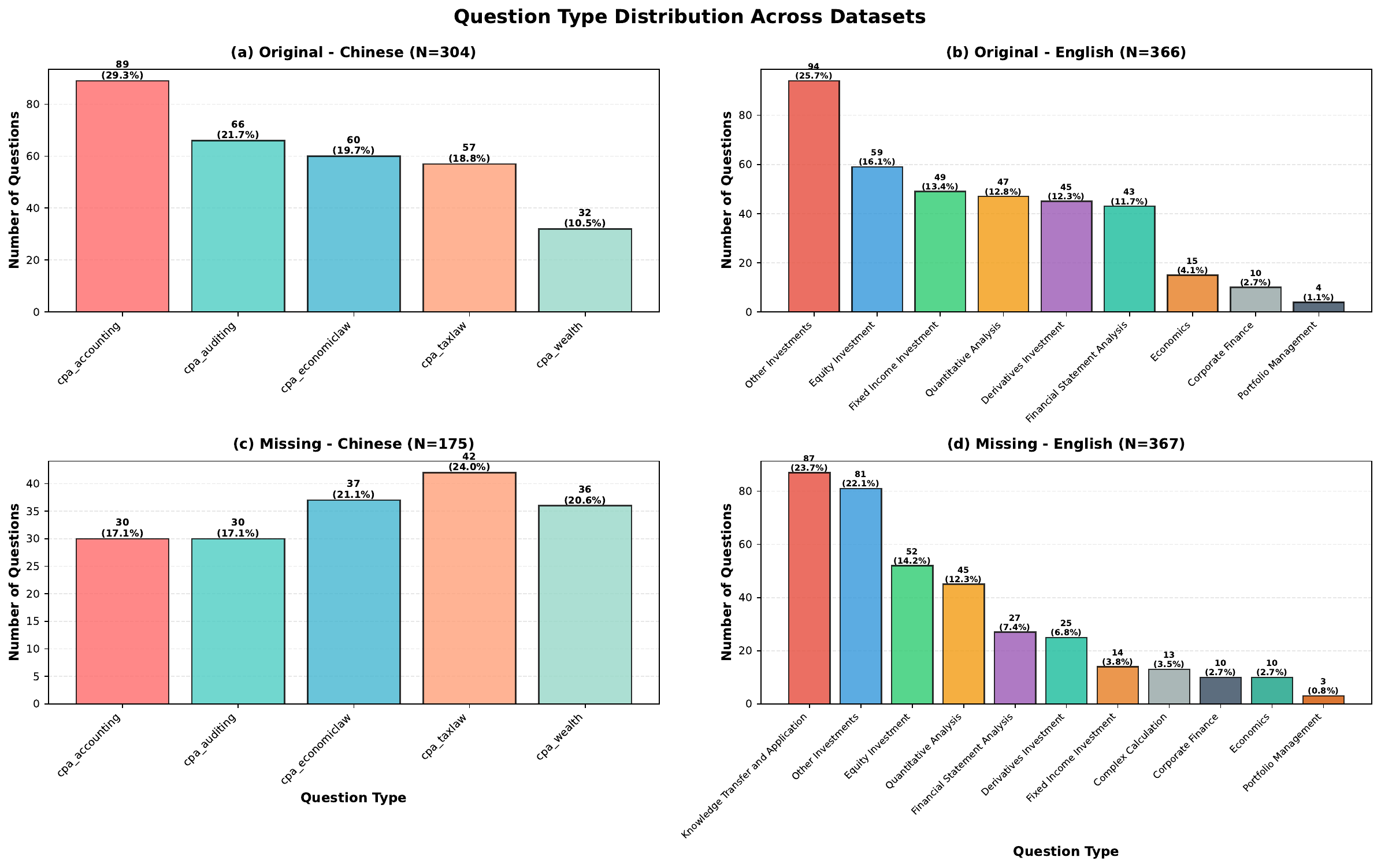}
  \caption{Question Type Distribution Across Datasets}
  \label{fig:distribution_combined_all}
\end{figure*}

\section{Statistical Analysis Methodology}
\label{app:statistics}

This appendix details the computational procedures and formulas 
used for all statistical analyses reported in Section~4.

\subsection{Difficulty Ranking Consistency (Spearman's $\rho$)}
\label{app:spearman}

\paragraph{Purpose.}
To measure whether different models exhibit consistent difficulty 
rankings across question classes.

\paragraph{Data Preparation.}
For each model and each question class, we compute the average 
accuracy as the percentage of correctly answered questions.
For model $m$ on question class $c$:
\begin{equation}
\text{Accuracy}(m, c) = \frac{\text{Number of correct answers}}
{\text{Total questions in class } c} \times 100\%
\end{equation}

\paragraph{Ranking Assignment.}
For each model $m$, we rank the question classes from hardest 
(rank 1) to easiest (rank $C$) based on accuracy:
\begin{equation}
R_m(c) = \text{rank of class } c \text{ for model } m
\end{equation}
where lower accuracy yields lower (harder) rank.

\paragraph{Pairwise Spearman Correlation.}
For each pair of models $(m_1, m_2)$, we compute Spearman's 
rank correlation coefficient:
\begin{equation}
\rho_{m_1, m_2} = 1 - \frac{6 \sum_{c=1}^{C} d_c^2}{C(C^2 - 1)}
\end{equation}
where $d_c = R_{m_1}(c) - R_{m_2}(c)$ is the difference in 
ranks for class $c$.

\paragraph{Average Correlation.}
The reported average Spearman's $\rho = 0.91$ is computed as:
\begin{equation}
\bar{\rho} = \frac{2}{M(M-1)} \sum_{m_1=1}^{M-1} 
\sum_{m_2=m_1+1}^{M} \rho_{m_1, m_2}
\end{equation}
where $M = 10$ models.

\subsection{Surface Feature Correlations}
\label{app:surface}

\paragraph{Question Length Analysis.}
For each question class $c$, we compute the average token count:
\begin{equation}
L_c = \frac{1}{N_c} \sum_{i=1}^{N_c} \text{tokens}(q_i)
\end{equation}
where $\text{tokens}(q_i)$ is the number of tokens in question $i$.

\paragraph{Independent Samples $t$-test.}
To test whether token length differs between the hardest 
(Knowledge Transfer, $L_{\text{KT}} = 127$) and easiest 
(Term Explanation, $L_{\text{TE}} = 134$) categories:
\begin{equation}
t = \frac{\bar{L}_{\text{KT}} - \bar{L}_{\text{TE}}}
{\sqrt{\frac{s_{\text{KT}}^2}{n_{\text{KT}}} + 
\frac{s_{\text{TE}}^2}{n_{\text{TE}}}}}
\end{equation}
where $s_c^2$ is the sample variance of token counts in class 
$c$, and $n_c$ is the sample size.

\paragraph{Numerical Density Correlation.}
We compute numerical density as:
\begin{equation}
D_c = \frac{1}{N_c} \sum_{i=1}^{N_c} 
\frac{\text{count of numbers in } q_i}{\text{tokens}(q_i)} 
\times 100
\end{equation}
Pearson correlation between $D_c$ and error rate 
$E_c = 100 - \text{Acc}_{\text{avg}}(c)$:
\begin{equation}
r_D = \frac{\sum_{c=1}^{C} (D_c - \bar{D})(E_c - \bar{E})}
{\sqrt{\sum_{c=1}^{C} (D_c - \bar{D})^2} 
\sqrt{\sum_{c=1}^{C} (E_c - \bar{E})^2}}
\end{equation}

\subsection{Information Integration Metrics}
\label{app:integration}

\paragraph{Feature Annotation.}
Each question was manually annotated by three expert annotators 
(Fleiss' $\kappa = 0.81$, substantial agreement) for:
\begin{itemize}
\item \textbf{Interacting Premises} ($P_i$): Number of distinct 
information pieces that must be combined.
\item \textbf{Cross-domain Dependencies} ($C_i$): Rated 1--5, 
whether reasoning spans multiple financial subdomains.
\item \textbf{Constraint Reconciliation} ($R_i$): Rated 1--5, 
complexity of reconciling conflicting requirements.
\end{itemize}

\paragraph{Aggregation.}
For each question class $c$:
\begin{align}
\bar{P}_c &= \frac{1}{N_c} \sum_{i=1}^{N_c} P_i \\
\bar{C}_c &= \frac{1}{N_c} \sum_{i=1}^{N_c} C_i \\
\bar{R}_c &= \frac{1}{N_c} \sum_{i=1}^{N_c} R_i
\end{align}

\paragraph{Pearson Correlation with Error Rate.}
For feature $X \in \{P, C, R\}$:
\begin{equation}
r_X = \frac{\sum_{c=1}^{C} (\bar{X}_c - \bar{\bar{X}})
(E_c - \bar{E})}{\sqrt{\sum_{c=1}^{C} 
(\bar{X}_c - \bar{\bar{X}})^2} 
\sqrt{\sum_{c=1}^{C} (E_c - \bar{E})^2}}
\end{equation}

\section{LLMs Implementation Details}

\subsection{LLMs Overview and Selection}
\label{app:model_details}

We set up a 5v5 comparison framework to make sure that the cross-paradigm examination was thorough and fair. This includes five cutting-edge general-purpose models that can be accessed through standard APIs:  GPT-5.1-mini, Gemini-2.5-Flash, Claude-3.5-Sonnet, DeepSeek-V3, and Qwen3-Max. These models were compared to five representative open-weight financial models that have been successfully used in our local inference environment: XuanYuan3-70B\citep{xuanyuan2.0}, Fin-R1-7B\citep{finr1}, CFGPT2-7B\citep{2023cfgpt}, DISC-FinLLM-13B\citep{2023discfinllm}, and FinGPT-7B\citep{2023fingpt}.

We started with a poll of possible candidates and then narrowed it down to this curated list based on strong standards of operational viability and technical transparency. Some intriguing models were left out because of administrative or proprietary issues. For example, InvestLM (HKUST)\citep{InvestLM} was still not available since the licensing approval process did not respond to multiple academic enquiries. Industrial systems like BloombergGPT\citep{bloomberggpt} and Baichuan4-Finance\citep{baichuan4} were also left out because their weights are strictly closed-source. Zhi Xiaobao, on the other hand, is part of a consumer-facing app that doesn't have a research-grade API. Also, ICBC and Hundsun's \footnote{\url{https://www.hs.net/lightgpt/}} internal infrastructure models can only be used in private banking settings, which makes it impossible to verify them independently.

Commercial API limits and architectural restrictions also required certain exclusions. HithinkGPT (Tonghuashun)\footnote{\url{https://aimiai.com/}} has a conversational interface, but its rigorous token-level quota of only 1,000 tokens with no institutional top-up option made it unsuitable for high-throughput reasoning audits. FinGPT-7B\citep{2023fingpt} was successfully deployed, but its tokeniser doesn't work well with Chinese characters; hence, Chinese tasks produce garbled outputs. 

To protect the integrity of the data and make sure the comparison was fair, FinGPT-7B's evaluation was limited to the English sub-benchmarks. Furthermore, across all question types in the Chinese CPA, all financial-specific models face collapse when encountering complex calculation problems, whether original or revised questions.

To directly address concerns about the coverage of recent reasoning-focused large models, we extend the core 5v5 suite with five additional reasoning-enhanced models spanning different families, scales, and training strategies: DeepSeek-R1\footnote{\url{https://huggingface.co/deepseek-ai/DeepSeek-R1}} and GPT-5.2-series, two frontier reasoning specialists evaluated via their official APIs; GPT-OSS-20B and GPT-OSS-120B\footnote{\url{https://huggingface.co/openai/gpt-oss-20b}; \url{https://huggingface.co/openai/gpt-oss-120b}}, OpenAI's open-weight Mixture-of-Experts reasoning models at 20B and 120B scale, deployed locally with native MXFP4 quantisation; and DianJin-R1-32B\footnote{\url{https://huggingface.co/DianJin/DianJin-R1-32B}}~\citep{DianJin-R1}, a Qwen2.5-base financial reasoning model trained with GRPO-style reinforcement learning. These five models were selected to cover the three strategies highlighted as underrepresented in the initial evaluation: reinforcement-learned reasoning (DeepSeek-R1, DianJin-R1-32B), frontier closed-source scaling (GPT-5.2-series), and open-weight reasoning at different parameter scales (GPT-OSS-20B/120B). Per-model configurations and deployment details are given in Appendix~\ref{app:xuanyuan3}--\ref{app:dianjin_r1}.


Table~\ref{tab:comprehensive_models} summarises representative commercial and financial LLMs, with all 15 evaluated models marked with $^{\ast}$. Looking ahead, we intend to broaden this research by including a wider spectrum of emerging financial LLMs, such as Fin-o1-8B\footnote{\url{https://huggingface.co/TheFinAI/Fin-o1-8B}}~\citep{fino1}, Finance-Llama\footnote{\url{https://huggingface.co/tarun7r/Finance-Llama-8B}}, Finance-Qwen\footnote{\url{https://huggingface.co/WiroAI/WiroAI-Finance-Qwen-1.5B}; \url{https://huggingface.co/WiroAI/WiroAI-Finance-Qwen-7B}}, and Fin-PRM\citep{finprm}. Future editions of this study will look into additional specialised open-source financial models as they become available in order to better benchmark the growing environment of financial AI.

Beyond static model evaluation, our future work will pivot towards autonomous financial intelligence. We aim to explore and develop frameworks similar to FinChain\citep{finchain}, focusing on structured financial reasoning chains, and FinAgent~\citep{finagent}, which emphasizes multi-agent collaboration for complex task automation. These directions will allow us to transition from benchmarking individual models to evaluating integrated financial AI systems in real-world operational environments.

\begin{table*}[ht]
\centering
\footnotesize
\setlength{\tabcolsep}{3pt}
\renewcommand{\arraystretch}{1.25}
\begin{tabularx}{\textwidth}{@{} l l l c >{\raggedright\arraybackslash}X >{\raggedright\arraybackslash}X @{} }
\toprule
\textbf{Category} & \textbf{Organization} & \textbf{Model} & \textbf{Size} & \textbf{Key Features / Notes} & \textbf{Source / Reference} \\
\midrule
\multirow{9}{*}{\shortstack[l]{General\\Purpose}} & \multirow{5}{*}{Closed-source}
  & \texttt{GPT-5.1-mini}$^{\ast}$        & --           & Frontier general-purpose performance  & openai/gpt-5.1-mini \\
& & \texttt{GPT-5.2-series}$^{\ast}$      & --           & Reasoning-enhanced frontier model     & openai/gpt-5.2 \\
& & \texttt{Claude-3.5-Sonnet}$^{\ast}$   & --           & Strong reasoning and stability        & anthropic/claude-3.5 \\
& & \texttt{Gemini-2.5-Flash}$^{\ast}$    & --           & High-efficiency reasoning model       & google/gemini-2.5 \\
& & \texttt{Qwen3-Max}$^{\ast}$           & $\sim$1T     & Proprietary MoE, leading multilingual & qwen/qwen3-max \\
\cmidrule{2-6}
& \multirow{4}{*}{Open-source}
  & \texttt{DeepSeek-V3}$^{\ast}$         & 37B/671B     & Advanced MoE reasoning focus          & deepseek/deepseek-v3 \\
& & \texttt{DeepSeek-R1}$^{\ast}$         & 37B/671B     & RL-trained reasoning specialist       & deepseek-ai/DeepSeek-R1 \\
& & \texttt{GPT-OSS-20B}$^{\ast}$         & 20B          & Open-weight MoE, native MXFP4         & openai/gpt-oss-20b \\
& & \texttt{GPT-OSS-120B}$^{\ast}$        & 120B         & Open-weight MoE, native MXFP4         & openai/gpt-oss-120b \\
\midrule
\multirow{7}{*}{Academic} & \multirow{7}{*}{Financial}
  & \texttt{FinGPT-7B}$^{\ast}$           & 7B   & Multi-source data curriculum          & AI4Finance/fingpt-7b \\
& & \texttt{InvestLM}                     & 65B  & Expert-level CFA-style reasoning      & HKUST/investlm \\
& & \texttt{DISC-FinLLM-13B}$^{\ast}$     & 13B  & Multi-turn advisory optimization      & Fudan/disc-finllm-13b \\
& & \texttt{CFGPT2-7B}$^{\ast}$           & 7B   & Knowledge graph integration           & Tongji/cfgpt2-7b \\
& & \texttt{Fin-R1-7B}$^{\ast}$           & 7B   & GRPO reinforcement learning           & SUFE/fin-r1-7b \\
& & \texttt{DianJin-R1-32B}$^{\ast}$      & 32B  & GRPO RL on Qwen2.5 base               & DianJin/DianJin-R1-32B \\
& & \texttt{FinMA-7B}                     & 7B   & Multi-task instruction tuning         & Pixiu/finma-7b \\
\midrule
\multirow{5}{*}{Industrial} & \multirow{5}{*}{Financial}
  & \texttt{BloombergGPT}                 & 50B  & Proprietary domain pre-training       & bloomberg/bloomberggpt \\
& & \texttt{XuanYuan3-70B}$^{\ast}$       & 70B  & Long-report synthesis focus           & duxiaoman/xuanyuan3-70b \\
& & \texttt{Baichuan4-Fin}                & --   & Professional logic alignment          & baichuan/baichuan4-fin \\
& & \texttt{HithinkGPT}                   & --   & Real-time API integration             & hithink/hithinkgpt \\
& & \texttt{Zhi Xiaobao}                  & --   & Consumer wealth management            & antgroup/zhixiaobao \\
\midrule
\multirow{2}{*}{\shortstack[l]{Infra-\\structure}} & \multirow{2}{*}{Financial}
  & \texttt{ICBC Zhiyong}                 & --   & Risk control and compliance           & icbc/zhiyong \\
& & \texttt{Light GPT}                    & --   & Securities trading backend            & lightgpt/lightgpt \\
\bottomrule
\end{tabularx}
\caption{Comprehensive Taxonomy of Large Language Models in the Financial Domain. Models marked with ($^{\ast}$) are those specifically evaluated in our empirical study, including 10 core-suite models and 5 extended reasoning-enhanced models added during revision.}
\label{tab:comprehensive_models}
\end{table*}

\subsection{Hardware Configuration}
\label{app:experiment_details}

The hardware assignment for each locally deployed model followed its memory footprint. XuanYuan3-70B, the largest dense model in our suite (70B parameters, $\sim$140~GB at FP16), was deployed on 2$\times$ NVIDIA H20 (96~GB each) with automatic tensor parallelism. DianJin-R1-32B (32B, Qwen2.5 base) fit on a single H20 (96~GB). The two GPT-OSS Mixture-of-Experts models, GPT-OSS-20B (21B total, $\sim$14~GB MXFP4) and GPT-OSS-120B (117B total, $\sim$63~GB MXFP4).They were each deployed on a single NVIDIA H100 (80~GB), since their native MXFP4 quantization and Triton MXFP4 kernel require a newer software environment than the rest of the suite (detailed below). DISC-FinLLM-13B (Baichuan-13B base) ran on a 32~GB vGPU instance. The remaining 7B-scale models (CFGPT2, FinGPT, Fin-R1) each fit on a single NVIDIA RTX 4090 (24~GB) at FP16.

In contrast, the seven frontier general-purpose and reasoning-specialist models, GPT-5.1-mini, GPT-5.2-series, Gemini-2.5-Flash, Claude-3.5-Sonnet, DeepSeek-V3, DeepSeek-R1, and Qwen3-Max. They were evaluated via their official APIs, which eliminated local GPU requirements while ensuring access to the latest model iterations.

All locally deployed experiments shared a unified software stack: PyTorch 2.1.0, Python 3.10 (Ubuntu 22.04), CUDA 12.1, and Transformers 4.38.2, with greedy decoding (\texttt{do\_sample=False}, \texttt{max\_new\_tokens=1024}, \texttt{repetition\_penalty=1.05}) to ensure deterministic outputs. The two GPT-OSS models are the only exception: because they rely on native MXFP4 quantization and the Triton \texttt{matmul\_ogs} MoE kernel unavailable in Transformers 4.38, they were run in a separate environment (Transformers $\geq$4.55, \texttt{triton} $\geq$3.4, and the \texttt{kernels} package) while retaining identical decoding settings.

\subsection{Prompts and Structured Output Protocol}
\label{app:prompts} 

All models are assessed under a unified protocol tailored to their specific deployment: locally implemented financial models utilize their native chat templates via \texttt{tokenizer. apply\_chat\_template()} to maintain strict alignment with original training paradigms, whereas general-purpose commercial models are accessed directly via standard API endpoints.

All experiments, which include the \textit{Original}, \textit{Revised}, and \textit{NOTA} settings, employ bilingual prompts mandating a structured JSON output with three fields: \texttt{reason}, \texttt{answer}, and \texttt{confidence}. To ensure robust data extraction and minimize evaluation bias, we develop a cascaded four-stage parsing strategy: the system first attempts to parse the entire output string using standard library decoders; if this fails, it proceeds to extract content within \texttt{json} or \texttt{} markdown tags; if the output remains unparsable, regular expressions are applied to identify the outermost balanced curly braces \texttt\{...\} to isolate the JSON object from potential conversational filler; finally, in cases of severe syntax corruption, we utilize field-level regex fallback to recover the \textit{reason}, \textit{answer}, and \textit{confidence} fields independently, thereby maximizing the recovery of partial but valid reasoning. For multiple-choice questions, extracted answers are normalized to uppercase and sorted alphabetically.

One major thing we saw when we looked at the five open-weight financial LLMs is the rise of a "formatting tax". This means that using strict JSON syntax usually lowers accuracy by 3\% to 4\% compared to free-form output. We hypothesize that the cognitive overhead of constraint satisfaction competes with the neural resources required for complex multi-step financial reasoning. However, we prioritize the structured protocol to ensure Epistemic Awareness can be explicitly audited through the reason field, decoupling logical deduction from linguistic fluency.

\begin{figure}[htbp]
    \centering
    \begin{tcolorbox}[eval_prompt, title=Chinese Prompt Template (CPA), left=5pt, right=5pt, top=5pt, bottom=5pt]
        \textbf{System:} \cn{您是一位金融问题解决专家。您将收到一道单选或多选题。请返回一个严格的 JSON 数据，包含三个键：}\texttt{"reason"} \cn{和} \texttt{"answer"} \cn{和} \texttt{"confidence"}\cn{。}

        \vspace{0.6em}
        \textbf{User:} \\
        \cn{问题：}\texttt{\{question\}}

        \vspace{0.4em}
        \cn{请严格按照以下格式输出 JSON 数据。请勿在 JSON 对象之外包含任何文本。}

        \vspace{0.4em}
        \texttt{\{} \\
        \quad \texttt{"reason": "}\cn{简要解释这些选项正确的原因。}\texttt{",} \\
        \quad \texttt{"answer": "}\cn{仅输出大写字母 (A--Z)。对于多选题，请返回按字母顺序排列并连接起来的字母集合（例如，ACD）。对于单选题，请返回字母，例如 A 或 B 或 C 或 D。}\texttt{",} \\
        \quad \texttt{"confidence": "}\cn{你对本题你所回答的答案的信心程度是多少，在 0--100 里选一个。}\texttt{"} \\
        \texttt{\}}
    \end{tcolorbox}
    \caption{The Chinese prompt template used for CPA-style evaluation. Applied consistently across all test variants.}
    \label{fig:prompt-zh}
\end{figure}

\begin{figure}[htbp]
    \centering
    \begin{tcolorbox}[eval_prompt, title=English Prompt Template (CFA), left=5pt, right=5pt, top=5pt, bottom=5pt]
        \textbf{System:} You are an expert in financial problem solving. You will be given a single- or multi-choice question. Please return STRICT JSON with three keys: \texttt{"reason"}, \texttt{"answer"} and \texttt{"confidence"}.

        \vspace{0.6em}
        \textbf{User:} \\
        Question: \texttt{\{question\}}

        \vspace{0.4em}
        Output JSON EXACTLY in the following format. Do NOT include any text outside the JSON object.

        \vspace{0.4em}
        \texttt{\{} \\
        \quad \texttt{"reason": "a concise explanation of why these options are correct.",} \\
        \quad \texttt{"answer": "output uppercase letters only (A--Z). For multi-choice, return the set of letters sorted alphabetically and concatenated (e.g., ACD). For single-choice, return letter like A or B or C or D.",} \\
        \quad \texttt{"confidence": "How confident are you in your answer to this question? Choose a score between 0 and 100."} \\
        \texttt{\}}
    \end{tcolorbox}
    \caption{The English prompt template used for CFA-style evaluation. Applied consistently across all test variants.}
    \label{fig:prompt-en}
\end{figure}

\subsection{XuanYuan3-70B Configuration}
\label{app:xuanyuan3}

\paragraph{Model Source.} \url{Duxiaoman-DI/Llama3-XuanYuan3-70B-Chat}\footnote{\url{https://huggingface.co/Duxiaoman-DI/Llama3-XuanYuan3-70B}}~\citep{xuanyuan2.0}


XuanYuan3-70B is a state-of-the-art Chinese large language financial model developed by Du Xiaoman. Building upon the Llama-3-70B architecture, it extends the foundational capabilities established in its predecessor, XuanYuan 2.0, through advanced post-training and financial-domain-specific instruction tuning. It is designed to handle sophisticated financial analysis, regulatory compliance, and cross-lingual financial reasoning with high parameter efficiency and domain expertise. We deploy XuanYuan3-70B using 2$\times$ NVIDIA H20 GPUs (96GB each) with automatic tensor parallelism, FP16 precision, greedy decoding, and \texttt{repetition\_penalty=1.05}.

\paragraph{Chat Template.} XuanYuan3-70B follows the native Llama-3 chat template, in which \texttt{system}, \texttt{user}, and \texttt{assistant} roles are explicitly encoded at the tokenizer level. We apply the template via \texttt{tokenizer.apply\_chat\_template()} with \texttt{add\_generation\_prompt=True} (Figure~\ref{fig:xuanyuan-template}).

\begin{figure}[h]
    \centering
    \tcbset{
        colframe=black!70,
        colback=gray!10,
        arc=2mm,
        boxrule=0.4pt,
        left=5pt,
        right=5pt,
        top=5pt,
        bottom=5pt
    }
    \begin{tcolorbox}
    \small
    \texttt{<|begin\_of\_text|>}\\
    \texttt{<|start\_header\_id|>system<|end\_header\_id|>}\\
    \\
    \texttt{\{system\}<|eot\_id|>}\\
    \texttt{<|start\_header\_id|>user<|end\_header\_id|>}\\
    \\
    \texttt{\{user\}<|eot\_id|>}\\
    \texttt{<|start\_header\_id|>assistant<|end\_header\_id|>}
    \end{tcolorbox}
    \caption{Llama-3 chat template used by XuanYuan3-70B.}
    \label{fig:xuanyuan-template}
\end{figure}

\paragraph{Detailed Results.}

Table~\ref{tab:xuanyuan_topic} demonstrates that the new strategy performs considerably better in all areas of finance. The model achieves nearly perfect results in Portfolio Management (100\%) and Derivatives Investment (92.00\%) on the English CFA track. The model shows a significant improvement in the Chinese CPA track, which is more crucial. Accuracy increased by 70.91\% in Auditing and 71.71\% in Economic Law. This shows that the optimization was effective in assisting the model in dealing with complex laws and regulations unique to distinct places.

\begin{table*}[!ht]
\centering
\small
\begin{tabular}{@{}lrrrc|lrrrc@{}}
\toprule
\textbf{Question Topic (CFA)} & \textit{Orig.} & \textit{Rev.} & $\Delta$ & \textit{n} & \textbf{Question Topic (CPA)} & \textit{Orig.} & \textit{Rev.} & $\Delta$ & \textit{n} \\
\midrule
Corporate Finance              & 60.00 & 90.00  & \cellcolor{posgreen}+30.00 & 10  & Accounting        &  4.49 & 63.33 & \cellcolor{posgreen}+58.84 & 89 \\
Derivatives Investment         & 53.33 & 92.00  & \cellcolor{posgreen}+38.67 & 45  & Auditing          & 25.76 & 96.67 & \cellcolor{posgreen}+70.91 & 66 \\
Economics                      & 53.33 & 80.00  & \cellcolor{posgreen}+26.67 & 15  & Economic Law      &  6.67 & 78.38 & \cellcolor{posgreen}+71.71 & 60 \\
Equity Investment              & 57.63 & 86.54  & \cellcolor{posgreen}+28.91 & 59  & Tax Law           & 15.79 & 66.67 & \cellcolor{posgreen}+50.88 & 57 \\
Financial Statement Analysis   & 65.12 & 88.89  & \cellcolor{posgreen}+23.77 & 43  & Wealth Management &  0.00 & 44.44 & \cellcolor{posgreen}+44.44 & 32 \\
Fixed Income Investment        & 42.86 & 78.57  & \cellcolor{posgreen}+35.71 & 49  &                   &       &       &        &    \\
Other Investments              & 62.77 & 86.42  & \cellcolor{posgreen}+23.65 & 94  &                   &       &       &        &    \\
Portfolio Management           & 75.00 & 100.00 & \cellcolor{posgreen}+25.00 &  4  &                   &       &       &        &    \\
Quantitative Analysis          & 65.96 & 86.67  & \cellcolor{posgreen}+20.71 & 47  &                   &       &       &        &    \\
\midrule
\textbf{Overall} & \textbf{58.47} & \textbf{84.74} & \textbf{+26.27} & \textbf{366} & \textbf{Overall} & \textbf{11.18} & \textbf{69.14} & \textbf{+57.96} & \textbf{304} \\
\bottomrule
\end{tabular}
\caption{Accuracy (\%) of XuanYuan3-70B by question topic. Left: English CFA (9 subjects). Right: Chinese CPA (5 subjects). \textit{n} denotes the number of original questions.}
\label{tab:xuanyuan_topic}
\end{table*}

Table \ref{tab:type_cpa_reordered} illustrates the proficiency gains across specific task categories. We observe a significant advancement in mathematical reasoning; within the English CFA, accuracy in Complex Calculation reaches 100.0\% under revised settings. Similarly, in the Chinese CPA, performance in Simple Calculation increases by more than 50.0\%. Most notably, XuanYuan3-70B effectively addresses the zero-shot performance collapse in CPA Knowledge Application, elevating accuracy from 0.00\% to 66.67\%. These results indicate that the model maintains professional-level proficiency across both conceptual frameworks and quantitative analysis.

\begin{table*}[!ht]
\centering
\small
\begin{tabular}{@{}lrrrr|rrrr@{}}
\toprule
 & \multicolumn{4}{c|}{\textbf{English CFA}} & \multicolumn{4}{c}{\textbf{Chinese CPA}} \\
\cmidrule(lr){2-5} \cmidrule(lr){6-9}
\textbf{Question Type} & \textit{Orig.} & \textit{Rev.} & $\Delta$ & \textit{n} & \textit{Orig.} & \textit{Rev.} & $\Delta$ & \textit{n} \\
\midrule
Conceptual Understanding  & 77.42 & 79.07  & \cellcolor{posgreen}+1.65  &  62 & 10.92 & 80.00 & \cellcolor{posgreen}+69.08 & 119 \\
Simple Calculation        & 36.67 & 82.12  & \cellcolor{posgreen}+45.46 &  30 &  9.76 & 64.15 & \cellcolor{posgreen}+54.39 &  41 \\
Complex Calculation       & 33.33 & 100.00 & \cellcolor{posgreen}+66.67 &  15 &  0.00 & --    & --                        &   2 \\
Comprehensive Judgment    & 62.38 & 88.37  & \cellcolor{posgreen}+26.00 & 101 & 10.34 & 71.72 & \cellcolor{posgreen}+61.37 &  87 \\
Knowledge Application     & 40.70 & 75.00  & \cellcolor{posgreen}+34.30 &  86 &  0.00 & 66.67 & \cellcolor{posgreen}+66.67 &   7 \\
Statistical Methods       & 71.43 & 93.62  & \cellcolor{posgreen}+22.19 &  70 & 16.67 & 60.00 & \cellcolor{posgreen}+43.33 &  48 \\
\midrule
\textbf{Overall} & \textbf{58.47} & \textbf{84.74} & \textbf{+26.27} & \textbf{364} & \textbf{11.18} & \textbf{69.14} & \textbf{+57.96} & \textbf{304} \\
\bottomrule
\end{tabular}
\caption{Accuracy (\%) of XuanYuan3-70B by question type across English CFA and Chinese CPA examinations.}
\label{tab:xuanyuan_qtype}
\end{table*}

\paragraph{Robustness Analysis under Adversarial Masking. }

The performance of XuanYuan3-70B within the NOTA-masked adversarial environment reveals a significant logical calibration challenge inherent to large-scale parameter models (Table~\ref{tab:answer-vs-reasoning-acc-gap}). The model exhibits a pronounced decoupling between reasoning trajectories and final label selection. Specifically, qualitative analysis shows that while the internal reasoning often converges on a specific numerical option, the model ultimately selects the ``None of the Above'' (NOTA) label due to surface-level pattern matching of the adversarial prompt structure. Furthermore, the model maintains exceptionally high confidence levels ($\mu > 90.0$) despite logical validity falling below 10\%, indicating a severe overconfidence bias. This suggests that while the model successfully identifies adversarial markers with high frequency in the English CFA (exceeding 80\%), it fails to ground its selection in a valid deductive process.


In summary, whilst XuanYuan3-70B utilizes a parameter count ten times that of Fin-R1-7B, it exhibits significantly lower logical consistency. This disparity suggests that the structural reasoning constraints established through reinforcement learning, specifically GRPO, represent a fundamental capability that cannot be replaced by the brute-force aesthetics of parameter expansion. These findings underscore that specialized reasoning paths are not emergent properties of scale alone but require targeted algorithmic constraints to ensure deductive integrity.

\subsection{Fin-R1-7B Configuration}
\label{app:finr1}

\paragraph{Model Source.} \texttt{SUFE-AIFLM-Lab/Fin-R1}\footnote{\url{https://huggingface.co/SUFE-AIFLM-Lab/Fin-R1}}~\citep{finr1}

Fin-R1 is developed by Shanghai University of Finance and Economics (SUFE), based on Qwen2.5-7B. It achieves state-of-the-art reasoning performance via GRPO-style reinforcement learning for complex financial problems. As of December 2025, it nearly matched the results of Xuanyuan3-70B. We use a single NVIDIA RTX 4090 (24GB) with FP16 precision, greedy decoding, and \texttt{repetition\_penalty=1.05}.

\paragraph{Chat Template.} Fin-R1 inherits the Qwen2.5 chat template, accessible via \texttt{tokenizer.apply\_chat\_template()}. We use the standard message format with system and user roles:

\begin{figure}[!ht]
    \centering
    \tcbset{
        colframe=black!70,
        colback=gray!10,
        arc=2mm,
        boxrule=0.4pt,
        left=5pt, right=5pt, top=5pt, bottom=5pt
    }
    \begin{tcolorbox}
    \small
\begin{verbatim}
messages = [
    {"role": "system", "content": system_prompt},
    {"role": "user", "content": user_prompt},
]
prompt = tokenizer.apply_chat_template(
    messages,
    tokenize=False,
    add_generation_prompt=True
)
\end{verbatim}
    \end{tcolorbox}
    \caption{Fin-R1 inference using official Qwen2.5 chat template.}
    \label{fig:finr1-code}
\end{figure}

Note: The official Fin-R1 repository recommends deployment via vLLM with an OpenAI-compatible API and a specific system prompt using \texttt{<think>...</think><answer>...</answer>} format for chain-of-thought reasoning. For consistency with our unified evaluation framework, we use direct inference with our standard JSON-output prompt instead.

\paragraph{Detailed Results.}

The evaluation results for Fin-R1-7B (Table~\ref{tab:finr1_topic} and \ref{tab:finr1_qtype}) present several significant insights. Despite possessing merely 7 billion parameters, Fin-R1-7B attains an overall accuracy of \textbf{83.65\%} in English CFA and \textbf{73.14\%} in Chinese CPA, closely approaching the performance of the significantly larger XuanYuan3-70B. This illustrates that GRPO-style reinforcement learning markedly improves the parameter efficacy of the base model, allowing it to internalise complex financial reasoning processes that generally necessitate higher-capacity architectures.

\begin{table*}[!ht]
\centering
\small
\begin{tabular}{@{}lrrrc|lrrrc@{}}
\toprule
\textbf{Question Topic (CFA)} & \textit{Orig.} & \textit{Rev.} & $\Delta$ & \textit{n} & \textbf{Question Topic (CPA)} & \textit{Orig.} & \textit{Rev.} & $\Delta$ & \textit{n} \\
\midrule
Corporate Finance              & 50.00 & 90.00  & \cellcolor{posgreen}+40.00 & 10  & Accounting        &  6.74 & 66.67 & \cellcolor{posgreen}+59.93 & 89 \\
Derivatives Investment         & 53.33 & 88.00  & \cellcolor{posgreen}+34.67 & 45  & Auditing          & 13.64 & 83.33 & \cellcolor{posgreen}+69.70 & 66 \\
Economics                      & 53.33 & 60.00  & \cellcolor{posgreen}+6.67  & 15  & Economic Law      &  6.67 & 91.89 & \cellcolor{posgreen}+85.22 & 60 \\
Equity Investment              & 62.71 & 86.54  & \cellcolor{posgreen}+23.83 & 59  & Tax Law           &  7.02 & 71.43 & \cellcolor{posgreen}+64.41 & 57 \\
Financial Statement Analysis   & 39.53 & 81.48  & \cellcolor{posgreen}+41.95 & 43  & Wealth Management &  0.00 & 52.78 & \cellcolor{posgreen}+52.78 & 32 \\
Fixed Income Investment        & 32.65 & 85.71  & \cellcolor{posgreen}+53.06 & 49  &                   &       &       &        &    \\
Other Investments              & 57.45 & 87.65  & \cellcolor{posgreen}+30.21 & 94  &                   &       &       &        &    \\
Portfolio Management           & 75.00 & 100.00 & \cellcolor{posgreen}+25.00 &  4  &                   &       &       &        &    \\
Quantitative Analysis          & 72.34 & 88.89  & \cellcolor{posgreen}+16.55 & 47  &                   &       &       &        &    \\
\midrule
\textbf{Overall} & \textbf{54.10} & \textbf{83.65} & \textbf{+29.55} & \textbf{366} & \textbf{Overall} & \textbf{7.57} & \textbf{73.14} & \textbf{+65.57} & \textbf{304} \\
\bottomrule
\end{tabular}
\caption{Accuracy (\%) of Fin-R1-7B by question topic. Left: English CFA (9 subjects). Right: Chinese CPA (5 subjects). \textit{n} denotes the number of original questions.}
\label{tab:finr1_topic}
\end{table*}

Fin-R1-7B exhibits a transformative accuracy surge within the Chinese CPA benchmark, rising from a near-random baseline of 7.57\% to 73.14\% ($\Delta$+65.57\%). Most notably, in Economic Law, the model achieves a peak accuracy of 91.89\%, surpassing several significantly larger architectures. This trajectory suggests that reasoning-oriented fine-tuning generalizes effectively across linguistic boundaries, even when navigating localized regulatory and legal frameworks.

As detailed in Table~\ref{tab:type_cpa_reordered}, Fin-R1-7B demonstrates substantial advancements in quantitative reasoning. Performance in Complex Calculation jumped from 26.67\% to 100.0\%, while Simple Calculation witnessed a gain exceeding 32\%. This improvement represents a hallmark of the reasoning model (R1) lineage, where the integration of structured chain-of-thought (CoT) processes—even when constrained to a JSON output format—significantly mitigates arithmetic and logical hallucinations.

The model maintains balanced proficiency across diverse cognitive tasks, specifically within Conceptual Understanding (90.00\% in CPA) and Statistical Methods (93.62\% in CFA). The consistent improvement ($\Delta$) across all task categories indicates that the reinforcement learning process does not merely optimize for specific keyword triggers but instead fosters a systematic enhancement in professional judgment and deductive integrity.

\begin{table*}[!ht]
\centering
\small
\begin{tabular}{@{}lrrrr|rrrr@{}}
\toprule
 & \multicolumn{4}{c|}{\textbf{English CFA}} & \multicolumn{4}{c}{\textbf{Chinese CPA}} \\
\cmidrule(lr){2-5} \cmidrule(lr){6-9}
\textbf{Question Type} & \textit{Orig.} & \textit{Rev.} & $\Delta$ & \textit{n} & \textit{Orig.} & \textit{Rev.} & $\Delta$ & \textit{n} \\
\midrule
Conceptual Understanding  & 69.35 & 76.74  & \cellcolor{posgreen}+7.39  &  62 &  7.56 & 90.00 & \cellcolor{posgreen}+82.44 & 119 \\
Simple Calculation        & 50.00 & 82.12  & \cellcolor{posgreen}+32.12 &  30 &  7.32 & 71.70 & \cellcolor{posgreen}+64.38 &  41 \\
Complex Calculation       & 26.67 & 100.00 & \cellcolor{posgreen}+73.33 &  15 &  0.00 & --    & --                        &   2 \\
Comprehensive Judgment    & 57.43 & 84.88  & \cellcolor{posgreen}+27.45 & 101 &  5.75 & 72.73 & \cellcolor{posgreen}+66.98 &  87 \\
Knowledge Application     & 38.37 & 75.00  & \cellcolor{posgreen}+36.63 &  86 & 14.29 & 66.67 & \cellcolor{posgreen}+52.38 &   7 \\
Statistical Methods       & 61.43 & 93.62  & \cellcolor{posgreen}+32.19 &  70 & 10.42 & 70.00 & \cellcolor{posgreen}+59.58 &  48 \\
\midrule
\textbf{Overall} & \textbf{54.10} & \textbf{83.65} & \textbf{+29.55} & \textbf{364} & \textbf{7.57} & \textbf{73.14} & \textbf{+65.57} & \textbf{304} \\
\bottomrule
\end{tabular}
\caption{Accuracy (\%) of Fin-R1-7B by question type across English CFA and Chinese CPA examinations.}
\label{tab:finr1_qtype}
\end{table*}

\paragraph{Robustness Analysis under Adversarial Masking. }

Fin-R1-7B demonstrates significant resilience in adversarial settings where the correct response is ``None of the Above'' (NOTA), achieving accuracies of 84.5\% in the CFA and 77.7\% in the CPA. Unlike general-purpose models that often exhibit a nearest-neighbor bias—the tendency to select the numerical option closest to an erroneous internal calculation—Fin-R1-7B maintains high reasoning consistency by actively rejecting plausible distractors through self-verification. The strong alignment between its high mean confidence ($\mu > 90.0$) and actual accuracy indicates a well-calibrated professional judgment likely fostered by GRPO reinforcement learning. This training process appears to effectively penalize logical shortcuts, enabling the model to identify specific flaws in distractor logic, such as the failure to account for tax-exempt income deductions. These findings highlight the parameter efficiency of specialized reasoning models, as the 7B architecture develops robust, logic-driven heuristics that rival the performance of much larger general architectures.


\subsection{CFGPT2-7B Configuration}
\label{app:cfgpt}

\paragraph{Model Source.} \texttt{TongjiFinLab/CFGPT2-7B}\footnote{\url{https://huggingface.co/TongjiFinLab/CFGPT2-7B}}~\citep{2023cfgpt}

CFGPT2 is a Chinese financial assistant developed by Tongji University, built upon the InternLM-7B base. It is pre-trained and fine-tuned on a massive corpus of Chinese financial data. We use a single NVIDIA RTX 4090 (24GB) with BF16 precision.

\paragraph{Chat Teamplate.} CFGPT2 uses the InternLM-style 
\texttt{model.chat()} method. The official example uses an empty 
\texttt{meta\_instruction}, but the interface supports passing 
system-level instructions:

\begin{figure}[h]
    \centering
    \tcbset{
  colframe=black!70,
  colback=gray!10,
  arc=2mm,
  boxrule=0.4pt,
  left=5pt,
  right=5pt,
  top=5pt,
  bottom=5pt}
    \begin{tcolorbox}
    \small
\begin{verbatim}
response, history = model.chat(
    tokenizer=tokenizer,
    query=user_prompt,
    history=[],
    max_new_tokens=1024,
    do_sample=False,
    repetition_penalty=1.1,
    meta_instruction=system_prompt
)
\end{verbatim}
    \end{tcolorbox}
    \caption{CFGPT2-7B inference code (InternLM-style).}
\end{figure}

\paragraph{Detailed Results.}

The evaluation of CFGPT2-7B (Table~\ref{tab:cfgpt2_topic} and \ref{tab:cfgpt2_qtype}) highlights the significant impact of [Optimization Method] on a domain-specific 7B model.

The original model (\textit{Orig.}) did almost nothing in the Chinese CPA track (2.96\%), which shows that it was very out of line with professional Chinese financial norms. But the amended version (\textit{Rev.}) jumped to 49.71\%, showing that fine-tuning can quickly bring out hidden financial information.
    
Like bigger reasoning models, CFGPT2-7B got a 100\% accuracy in English \textit{Complex Calculation} after revision. The increases in \textit{Simple Calculation} were more than 38\% and 44\% across both benchmarks.

\begin{table*}[t!]
\centering
\small
\begin{tabular}{@{}lrrrc|lrrrc@{}}
\toprule
\textbf{Question Topic (CFA)} & \textit{Orig.} & \textit{Rev.} & $\Delta$ & \textit{n} & \textbf{Question Topic (CPA)} & \textit{Orig.} & \textit{Rev.} & $\Delta$ & \textit{n} \\
\midrule
Corporate Finance              & 60.00 & 90.00  & \cellcolor{posgreen}+30.00 & 10  & Accounting        &  2.25 & 60.00 & \cellcolor{posgreen}+57.75 & 89 \\
Derivatives Investment         & 31.11 & 76.00  & \cellcolor{posgreen}+44.89 & 45  & Auditing          &  7.58 & 60.00 & \cellcolor{posgreen}+52.42 & 66 \\
Economics                      & 53.33 & 60.00  & \cellcolor{posgreen}+6.67  & 15  & Economic Law      &  1.67 & 45.95 & \cellcolor{posgreen}+44.28 & 60 \\
Equity Investment              & 59.32 & 78.85  & \cellcolor{posgreen}+19.53 & 59  & Tax Law           &  1.75 & 40.48 & \cellcolor{posgreen}+38.73 & 57 \\
Financial Statement Analysis   & 53.49 & 66.67  & \cellcolor{posgreen}+13.18 & 43  & Wealth Management &  0.00 & 47.22 & \cellcolor{posgreen}+47.22 & 32 \\
Fixed Income Investment        & 36.73 & 57.14  & \cellcolor{posgreen}+20.41 & 49  &                   &       &       &        &    \\
Other Investments              & 50.00 & 77.78  & \cellcolor{posgreen}+27.78 & 94  &                   &       &       &        &    \\
Portfolio Management           & 75.00 & 100.00 & \cellcolor{posgreen}+25.00 &  4  &                   &       &       &        &    \\
Quantitative Analysis          & 42.55 & 73.33  & \cellcolor{posgreen}+30.78 & 47  &                   &       &       &        &    \\
\midrule
\textbf{Overall} & \textbf{47.54} & \textbf{73.57} & \textbf{+26.03} & \textbf{366} & \textbf{Overall} & \textbf{2.96} & \textbf{49.71} & \textbf{+46.75} & \textbf{304} \\
\bottomrule
\end{tabular}
\caption{Accuracy (\%) of CFGPT2-7B by question topic. Left: English CFA (9 subjects). Right: Chinese CPA (5 subjects). \textit{n} denotes the number of original questions.}
\label{tab:cfgpt2_topic}
\end{table*}

The model is quite good in CFA's \textit{Corporate Finance} (\textbf{90.00\%}) and \textit{Portfolio Management} (100\%). In CPA, it scored 80.00\% in \textit{Statistical Methods}, making it a great candidate for quantitative financial jobs even though it has fewer parameters.

\begin{table*}[t!]
\centering
\small
\begin{tabular}{@{}lrrrr|rrrr@{}}
\toprule
 & \multicolumn{4}{c|}{\textbf{English CFA}} & \multicolumn{4}{c}{\textbf{Chinese CPA}} \\
\cmidrule(lr){2-5} \cmidrule(lr){6-9}
\textbf{Question Type} & \textit{Orig.} & \textit{Rev.} & $\Delta$ & \textit{n} & \textit{Orig.} & \textit{Rev.} & $\Delta$ & \textit{n} \\
\midrule
Conceptual Understanding  & 56.45 & 76.74  & \cellcolor{posgreen}+20.29 &  62 &  2.52 & 40.00 & \cellcolor{posgreen}+37.48 & 119 \\
Simple Calculation        & 30.00 & 68.16  & \cellcolor{posgreen}+38.16 &  30 &  4.88 & 49.06 & \cellcolor{posgreen}+44.18 &  41 \\
Complex Calculation       & 40.00 & 100.00 & \cellcolor{posgreen}+60.00 &  15 &  0.00 & --    & --                        &   2 \\
Comprehensive Judgment    & 60.40 & 80.23  & \cellcolor{posgreen}+19.83 & 101 &  2.30 & 49.49 & \cellcolor{posgreen}+47.19 &  87 \\
Knowledge Application     & 31.40 & 75.00  & \cellcolor{posgreen}+43.60 &  86 &  0.00 &  0.00 & 0.00                      &   7 \\
Statistical Methods       & 48.57 & 76.60  & \cellcolor{posgreen}+28.03 &  70 &  4.17 & 80.00 & \cellcolor{posgreen}+75.83 &  48 \\
\midrule
\textbf{Overall} & \textbf{47.54} & \textbf{73.57} & \textbf{+26.03} & \textbf{364} & \textbf{2.96} & \textbf{49.71} & \textbf{+46.75} & \textbf{304} \\
\bottomrule
\end{tabular}
\caption{Accuracy (\%) of CFGPT2-7B by question type across English CFA and Chinese CPA examinations.}
\label{tab:cfgpt2_qtype}
\end{table*}

\paragraph{Robustness Analysis under Adversarial Masking. }

CFGPT2-7B exhibits a critical \textit{reasoning-selection gap}, where its moderate \textit{Answer Accuracy} (75.2\% in CFA; 54.9\% in CPA) is starkly contradicted by its near-complete collapse in \textit{Reasoning Accuracy} ($\sim$10\%). This suggests that the model's correct outputs are largely driven by stochastic output heuristics rather than deterministic derivation. Unlike Fin-R1-7B, CFGPT2-7B lacks a functional verification closed-loop (i.e., the logical gate: ``Calculated Result $\neq$ Candidates $\rightarrow$ Select NOTA''), often leading to a correct label despite a fallacious reasoning path.

The performance of CFGPT2-7B is characterized by unstable hallucination-driven hits. Qualitative traces indicate that even when the model correctly identifies the NOTA mask, its CoT frequently contains factual distortions or logical non-sequiturs. This implies that financial domain knowledge has not been internalized into a robust reasoning faculty but remains at the level of probabilistic surface matching.

Despite its logical invalidity, CFGPT2-7B maintains an average confidence score of $\mu > 81.0$. This severe miscalibration represents a significant risk in financial applications; the model exhibits a ``blind certainty,'' providing authoritative yet erroneous justifications.


\subsection{DISC-FinLLM-13B Configuration}
\label{app:disc-finllm}

\paragraph{Model Source.} \texttt{Go4miii/DISC-FinLLM}\footnote{\url{https://huggingface.co/Go4miii/DISC-FinLLM}}~\citep{2023discfinllm}

DISC-FinLLM is a specialised financial large language model fine-tuned on the Baichuan-13B base. It is worth noting that DISC-FinLLM is primarily engineered for multi-turn financial consulting dialogues rather than single-turn MCQ tasks. We use a virtual GPU (vGPU-32GB) with FP16 precision and greedy decoding.

\paragraph{Chat Template.} DISC-FinLLM uses the Baichuan-style \texttt{model.chat()} method, which does not support a separate system role. To incorporate our evaluation instructions, we concatenate the system prompt with the user query into a single user message:

\begin{figure}[h]
    \centering
    \tcbset{
        colframe=black!70,
        colback=gray!10,
        arc=2mm,
        boxrule=0.4pt,
        left=5pt, right=5pt, top=5pt, bottom=5pt
    }
    \begin{tcolorbox}
    \small
\begin{verbatim}
full_content = f"{system_prompt}\n\n
{user_content}"

messages = [{"role": "user", "content": 
full_content}]

response = model.chat(tokenizer, messages)
\end{verbatim}
    \end{tcolorbox}
    \caption{DISC-FinLLM inference with merged system prompt.}
    \label{fig:disc-finllm-code}
\end{figure}

\paragraph{Detailed Results.}

The evaluation results for DISC-FinLLM (Table~\ref{tab:disc_topic} and \ref{tab:disc_qtype}) provide insights into the cross-task generalisation of dialogue-specialised financial models

As a model primarily optimised for multi-turn financial consulting, DISC-FinLLM exhibited relatively low baseline accuracy in both the English CFA (13.11\%) and Chinese CPA (4.61\%) tracks. This suggests that the instruction-tuning for conversational equity does not naturally translate into the rigorous logical deduction required for professional MCQs. 

Although the revised version (\textit{Rev.}) showed improvement, climbing to 19.89\% in CFA and 19.43\% in CPA. The gains are significantly more modest compared to reasoning-dense models like Fin-R1. Notably, in several CFA subjects such as \textit{Economics} and \textit{Equity Investment}, the model even experienced performance regression ($\Delta < 0$), indicating a potential instruction interference where revised prompts might conflict with its conversational priors.

The question-type breakdown reveals that DISC-FinLLM-13B struggles with Statistical Methods and Quantitative Analysis, often scoring near or below the expected value of random guessing. While it showed a minor breakthrough in CPA \textit{Conceptual Understanding} (rising from 4.20\% to 30.00\%), its overall performance trajectory confirms that without specific reinforcement learning for reasoning paths, dialogue-centric models remain insufficient for autonomous high-stakes financial examination tasks.

\begin{table*}[t!]
\centering
\small
\begin{tabular}{@{}lrrrc|lrrrc@{}}
\toprule
\textbf{Question Topic (CFA)} & \textit{Orig.} & \textit{Rev.} & $\Delta$ & \textit{n} & \textbf{Question Topic (CPA)} & \textit{Orig.} & \textit{Rev.} & $\Delta$ & \textit{n} \\
\midrule
Corporate Finance              & 30.00 & 40.00  & \cellcolor{posgreen}+10.00 & 10  & Accounting        &  4.49 & 26.67 & \cellcolor{posgreen}+22.18 & 89 \\
Derivatives Investment         & 20.00 & 28.00  & \cellcolor{posgreen}+8.00  & 45  & Auditing          &  4.55 & 26.67 & \cellcolor{posgreen}+22.12 & 66 \\
Economics                      & 13.33 &  0.00  & \cellcolor{negred}--13.33  & 15  & Economic Law      &  6.67 & 18.92 & \cellcolor{posgreen}+12.25 & 60 \\
Equity Investment              & 20.34 & 13.46  & \cellcolor{negred}--6.88   & 59  & Tax Law           &  5.26 & 14.29 & \cellcolor{posgreen}+9.03  & 57 \\
Financial Statement Analysis   & 11.63 & 22.22  & \cellcolor{posgreen}+10.59 & 43  & Wealth Management &  0.00 & 13.89 & \cellcolor{posgreen}+13.89 & 32 \\
Fixed Income Investment        & 14.29 & 14.29  & 0.00                       & 49  &                   &       &       &        &    \\
Other Investments              & 10.64 & 23.46  & \cellcolor{posgreen}+12.82 & 94  &                   &       &       &        &    \\
Portfolio Management           &  0.00 & 33.33  & \cellcolor{posgreen}+33.33 &  4  &                   &       &       &        &    \\
Quantitative Analysis          &  0.00 &  6.67  & \cellcolor{posgreen}+6.67  & 47  &                   &       &       &        &    \\
\midrule
\textbf{Overall} & \textbf{13.11} & \textbf{19.89} & \textbf{+6.78} & \textbf{366} & \textbf{Overall} & \textbf{4.61} & \textbf{19.43} & \textbf{+14.82} & \textbf{304} \\
\bottomrule
\end{tabular}
\caption{Accuracy (\%) of DISC-FinLLM-13B by question topic. Left: English CFA (9 subjects). Right: Chinese CPA (5 subjects). \textit{n} denotes the number of original questions.}
\label{tab:disc_topic}
\end{table*}

\begin{table*}[t!]
\centering
\small
\begin{tabular}{@{}lrrrr|rrrr@{}}
\toprule
 & \multicolumn{4}{c|}{\textbf{English CFA}} & \multicolumn{4}{c}{\textbf{Chinese CPA}} \\
\cmidrule(lr){2-5} \cmidrule(lr){6-9}
\textbf{Question Type} & \textit{Orig.} & \textit{Rev.} & $\Delta$ & \textit{n} & \textit{Orig.} & \textit{Rev.} & $\Delta$ & \textit{n} \\
\midrule
Conceptual Understanding  &  8.06 & 18.60  & \cellcolor{posgreen}+10.54 &  62 &  4.20 & 30.00 & \cellcolor{posgreen}+25.80 & 119 \\
Simple Calculation        & 23.33 & 20.67  & \cellcolor{negred}--2.66   &  30 &  4.88 & 15.09 & \cellcolor{posgreen}+10.21 &  41 \\
Complex Calculation       &  6.67 & 25.00  & \cellcolor{posgreen}+18.33 &  15 &  0.00 & --    & --                        &   2 \\
Comprehensive Judgment    & 22.77 & 22.09  & \cellcolor{negred}--0.68   & 101 &  5.75 & 20.20 & \cellcolor{posgreen}+14.45 &  87 \\
Knowledge Application     & 10.47 & 12.50  & \cellcolor{posgreen}+2.03  &  86 & 14.29 &  0.00 & \cellcolor{negred}--14.29  &   7 \\
Statistical Methods       &  2.86 & 14.89  & \cellcolor{posgreen}+12.03 &  70 &  2.08 & 30.00 & \cellcolor{posgreen}+27.92 &  48 \\
\midrule
\textbf{Overall} & \textbf{13.11} & \textbf{19.89} & \textbf{+6.78} & \textbf{364} & \textbf{4.61} & \textbf{19.43} & \textbf{+14.82} & \textbf{304} \\
\bottomrule
\end{tabular}
\caption{Accuracy (\%) of DISC-FinLLM-13B by question type across English CFA and Chinese CPA examinations.}
\label{tab:disc_qtype}
\end{table*}

\paragraph{Robustness Analysis under Adversarial Masking. }

The erratic performance and diminished confidence ratings, particularly in English CFA, suggest that DISC-FinLLM-13B is acutely susceptible to hostile prompt alterations, lacking the ability to generalise outside its conversational tuning distribution. The deterioration of reasoning accuracy indicates that the model lacks the structural logic necessary to recognise absent information, instead depending on language templates that are readily undermined in NOTA situations.


\subsection{FinGPT Configuration}
\label{app:fingpt}

\paragraph{Model Source.}
\textit{AI4Finance-Foundation/FinGPT}~\citep{2023fingpt}.\footnote{Base
and LoRA adapter: \url{https://huggingface.co/Qwen/Qwen-7B} and
\url{https://huggingface.co/FinGPT/fingpt-mt_qwen-7b_lora} respectively.}
We utilised the \textbf{Qwen-7B-based variant} of FinGPT to evaluate the
English CFA tracks (Original and Revised). Among the FinGPT variant
family (Llama-2, Falcon, Bloom, MPT, ChatGLM2, Qwen), we selected the
Qwen variant because the Qwen-7B tokeniser natively handles Chinese
characters, whereas Llama-2- and Falcon-based variants corrupt Chinese
inputs at the tokenisation stage. Preliminary experiments, however,
revealed a second, subtler limitation specific to this variant, which
we term \textbf{Reasoning Dissociation}: the LoRA adapter, fine-tuned
primarily on English financial sentiment data, interferes with the
Qwen-7B base's generative consistency in zero-shot professional
reasoning. On English CFA items we observed fragmented, largely
uninformative outputs (e.g., truncated phrases such as
\texttt{pass\_through} or \texttt{Acorp} in place of structured
reasoning), which precluded meaningful accuracy gains from the
adapter. To preserve evaluation integrity across our model suite, we
therefore limited FinGPT's evaluation to the English CFA sub-benchmark.

\paragraph{Chat Template.} FinGPT utilises an explicit
\texttt{Instruction/Input/Answer} format, inherited from its multi-task
fine-tuning paradigm in which the non-chat base checkpoint is adapted
via LoRA directly on plain-text instructions. To maintain consistency
with this training distribution, we mapped the evaluation system
prompt to the \textit{Instruction} field and the user prompt to the
\textit{Input} field (Figure~\ref{fig:fingpt-code}). Critically, we do
\textit{not} invoke \texttt{tokenizer.apply\_chat\_template()} here,
since the Qwen-7B \textit{base} checkpoint (as distinct from
\textit{Qwen-7B-Chat}) does not ship with a chat template, and FinGPT's
adapter was not trained under any chat-formatted input distribution.

\begin{figure}[!t]
    \centering
    \tcbset{colframe=black!70, colback=gray!10, arc=2mm, boxrule=0.4pt,
            left=5pt, right=5pt, top=5pt, bottom=5pt}
    \begin{tcolorbox}
    \small
    \begin{verbatim}
# FinGPT (Qwen-7B base + LoRA) inference logic
prompt = f"Instruction: {system_prompt}\n
Input: {user_prompt}\n
Answer: "
    \end{verbatim}
    \end{tcolorbox}
    \caption{FinGPT (Qwen-7B base + LoRA) inference template for
    English CFA tasks. All FinGPT variants share this flat
    \texttt{Instruction/Input/Answer} format; no chat template is
    applied, since FinGPT's LoRA was fine-tuned directly on the
    non-chat base checkpoint.}
    \label{fig:fingpt-code}
\end{figure}


\paragraph{Detailed Results.}
As shown in \tabref{tab:fingpt_combined}, FinGPT-7B achieved an overall accuracy of only \textbf{2.19\%} (Original) and 3.54\% (Revised), significantly below random guessing. Qualitative inspection revealed that the model frequently outputs sentiment markers (e.g., ``positive/negative'') or jumbled characters instead of logical steps. This confirms that its fine-tuning distribution—highly specialized for sentiment classification—severely compromises its capacity for multi-step financial deduction.

\begin{table}[t!]
    \centering
    \small
    
        \begin{tabular}{@{}lrrrc@{}}
            \toprule
            \textbf{Topic} & \textit{Orig.} & \textit{Rev.} & $\Delta$ & \textit{n} \\
            \midrule
            Corp. Fin.   & 0.00 & 0.00  & 0.00                       & 10 \\
            Derivatives  & 6.67 & 4.00  & \cellcolor{negred}--2.67   & 45 \\
            Economics    & 0.00 & 0.00  & 0.00                       & 15 \\
            Equity       & 1.69 & 3.85  & \cellcolor{posgreen}+2.16  & 59 \\
            FSA          & 2.33 & 3.70  & \cellcolor{posgreen}+1.37  & 43 \\
            Fixed Inc.   & 0.00 & 0.00  & 0.00                       & 49 \\
            Other Inv.   & 3.19 & 3.70  & \cellcolor{posgreen}+0.51  & 94 \\
            Portfolio    & 0.00 & 33.33 & \cellcolor{posgreen}+33.33 & 4  \\
            Quant.       & 0.00 & 2.22  & \cellcolor{posgreen}+2.22  & 47 \\
            \midrule
            \textbf{Overall} & \textbf{2.19} & \textbf{3.54} & \textbf{+1.35} & \textbf{366} \\
    
    
            \midrule
            \textbf{Type} & \textit{Orig.} & \textit{Rev.} & $\Delta$ & \textit{n} \\
            \midrule
            Conceptual     & 1.61 & 6.98 & \cellcolor{posgreen}+5.37 & 62  \\
            Simple Calc.   & 6.67 & 2.79 & \cellcolor{negred}--3.88  & 30  \\
            Complex Calc.  & 0.00 & 0.00 & 0.00                      & 15  \\
            Comp. Judgment & 1.98 & 4.65 & \cellcolor{posgreen}+2.67 & 101 \\
            Know. App.     & 0.00 & 0.00 & 0.00                      & 86  \\
            Stat. Methods  & 4.29 & 2.13 & \cellcolor{negred}--2.16  & 70  \\
            \midrule
            \textbf{Overall} & \textbf{2.19} & \textbf{3.54} & \textbf{+1.35} & \textbf{364} \\
            \bottomrule
        \end{tabular}
    
    \caption{Performance of FinGPT-7B on English CFA Benchmark.}
    \label{tab:fingpt_combined}
\end{table}

\paragraph{Robustness Analysis under Adversarial Masking.}
FinGPT-7B undergoes a total functional collapse in NOTA-masked scenarios (Table~\ref{tab:answer-vs-reasoning-acc-gap}). The near-zero reasoning accuracy and stagnant confidence score ($\sim$50.0) reveal an absolute lack of \textit{logical plasticity}. Rather than identifying missing information, the model reverts to fixed templates, proving that sentiment-optimized models remain inherently fragile in adversarial professional environments.

\subsection{GPT-OSS-20B and GPT-OSS-120B Configuration}
\label{app:gpt_oss}

\paragraph{Model Sources.} \texttt{openai/gpt-oss-20b}\footnote{\url{https://huggingface.co/openai/gpt-oss-20b}} and \texttt{openai/gpt-oss-120b}\footnote{\url{https://huggingface.co/openai/gpt-oss-120b}}

The two GPT-OSS models are open-weight reasoning Mixture-of-Experts (MoE) models released by OpenAI under the Apache 2.0 licence. GPT-OSS-20B has 21B total parameters with approximately 3.6B active parameters per forward pass (32 experts, top-4 routing). GPT-OSS-120B has 117B total parameters with approximately 5.1B active parameters (128 experts, top-4 routing). Both ship with native MXFP4 quantisation of the MoE weights and BF16 storage for all other tensors. The MXFP4 footprint is approximately 14\,GB for the 20B variant and 63\,GB for the 120B variant, reduced from $\sim$240\,GB at BF16. Both models were deployed on a single NVIDIA H100 (80\,GB), which motivates their joint treatment in this subsection. Including both variants lets us isolate the effect of parameter scaling under a fixed training paradigm and chat template.

\paragraph{Chat Template.}
Both models use the built-in Harmony response format, accessed through the HuggingFace \texttt{transformers} pipeline at \texttt{transformers}$\geq$4.55, together with the Triton \texttt{matmul\_ogs} MXFP4 kernel (\texttt{triton}$\geq$3.4, \texttt{kernels}). The chat template and decoding settings are identical across the two models:

\begin{figure}[h]
    \centering
    \tcbset{
  colframe=black!70,
  colback=gray!10,
  arc=2mm,
  boxrule=0.4pt,
  left=5pt,
  right=5pt,
  top=5pt,
  bottom=5pt}
    \begin{tcolorbox}
    \small
\begin{verbatim}
messages = [
    {"role": "system", "content": system_prompt},
    {"role": "user",   "content": user_prompt},
]
prompt = tokenizer.apply_chat_template(
    messages, tokenize=False,
    add_generation_prompt=True)

# greedy decoding, consistent with the suite
outputs = model.generate(
    **tokenizer(prompt, return_tensors="pt"),
    do_sample=False,
    max_new_tokens=1024,
    repetition_penalty=1.05,
)
\end{verbatim}
    \end{tcolorbox}
    \caption{GPT-OSS inference code (Harmony chat template). The same code applies to both GPT-OSS-20B and GPT-OSS-120B.}
\end{figure}

\paragraph{Detailed Results: GPT-OSS-20B.}

Table~\ref{tab:gptoss20b_topic} shows a profile typical of a general-purpose reasoning model rather than a finance-tuned one. English CFA \textit{Orig.}\ accuracy is already 78.22\% without any financial fine-tuning, and \textit{Rev.}\ accuracy reaches 88.58\% (+10.36\%). Chinese CPA reveals a sharper pattern. \textit{Orig.}\ accuracy collapses to 12.00\%, the same dead-zone signature observed in Chinese-financial-specialized models, yet \textit{Rev.}\ accuracy rises to 77.63\% ({+65.63\%}). A 65-point Chinese jump in the absence of Chinese-CPA training data is consistent with our main-paper account of \textit{implicit assumption alignment}: the model fills the missing premise with a plausible default and commits to it.

\begin{table*}[t!]
\centering
\small
\begin{tabular}{@{}lrrrc|lrrrc@{}}
\toprule
\textbf{Question Topic (CFA)} & \textit{Orig.} & \textit{Rev.} & $\Delta$ & \textit{n} & \textbf{Question Topic (CPA)} & \textit{Orig.} & \textit{Rev.} & $\Delta$ & \textit{n} \\
\midrule
Corporate Finance              &  80.00 &  90.00 & \cellcolor{posgreen}+10.00 & 10  & Accounting        & 10.00 & 45.83 & \cellcolor{posgreen}+35.83 & 89 \\
Derivatives Investment         &  79.41 &  91.67 & \cellcolor{posgreen}+12.25 & 45  & Auditing          & 19.30 & 93.33 & \cellcolor{posgreen}+74.04 & 66 \\
Economics                      &  90.91 &  90.00 & \cellcolor{negred}$-$0.91  & 15  & Economic Law      &  4.88 & 81.25 & \cellcolor{posgreen}+76.37 & 60 \\
Equity Investment              &  83.02 &  88.46 & \cellcolor{posgreen}+5.44  & 59  & Tax Law           & 10.26 & 94.59 & \cellcolor{posgreen}+84.34 & 57 \\
Financial Statement Analysis   &  76.19 &  85.19 & \cellcolor{posgreen}+8.99  & 43  & Wealth Management & 14.29 & 62.07 & \cellcolor{posgreen}+47.78 & 32 \\
Fixed Income Investment        &  63.41 &  78.57 & \cellcolor{posgreen}+15.16 & 49  &                   &       &       &        &    \\
Other Investments              &  76.19 &  88.89 & \cellcolor{posgreen}+12.70 & 94  &                   &       &       &        &    \\
Portfolio Management           & 100.00 & 100.00 & +0.00                      &  4  &                   &       &       &        &    \\
Quantitative Analysis          &  85.11 &  88.89 & \cellcolor{posgreen}+3.78  & 47  &                   &       &       &        &    \\
\midrule
\textbf{Overall} & \textbf{78.22} & \textbf{88.58} & \textbf{+10.36} & \textbf{366} & \textbf{Overall} & \textbf{12.00} & \textbf{77.63} & \textbf{+65.63} & \textbf{304} \\
\bottomrule
\end{tabular}
\caption{Accuracy (\%) of GPT-OSS-20B by question topic. Left: English CFA (9 subjects). Right: Chinese CPA (5 subjects). \textit{n} denotes the number of original questions.}
\label{tab:gptoss20b_topic}
\end{table*}

English performance is uniformly strong, with 100\% on \textit{Portfolio Management} and above 85\% on \textit{Quantitative Analysis}, \textit{Equity Investment}, and \textit{Economics}. The weakest topic is \textit{Fixed Income Investment} at 63.41\%. On Chinese CPA, \textit{Rev.}\ accuracy exceeds 80\% on the three rule-prescriptive topics (\textit{Auditing}, \textit{Economic Law}, \textit{Tax Law}) but remains below 65\% on \textit{Wealth Management}, the cross-product reasoning topic.

\begin{table*}[t!]
\centering
\small
\begin{tabular}{@{}lrrrr|rrrr@{}}
\toprule
 & \multicolumn{4}{c|}{\textbf{English CFA}} & \multicolumn{4}{c}{\textbf{Chinese CPA}} \\
\cmidrule(lr){2-5} \cmidrule(lr){6-9}
\textbf{Question Type} & \textit{Orig.} & \textit{Rev.} & $\Delta$ & \textit{n} & \textit{Orig.} & \textit{Rev.} & $\Delta$ & \textit{n} \\
\midrule
Conceptual Understanding  & 85.25 &  83.72 & \cellcolor{negred}$-$1.52  &  62 & 12.12 &  90.00 & \cellcolor{posgreen}+77.88 & 119 \\
Simple Calculation        & 86.36 &  88.37 & \cellcolor{posgreen}+2.01  &  30 &  4.76 &  65.00 & \cellcolor{posgreen}+60.24 &  41 \\
Complex Calculation       & 75.00 & 100.00 & \cellcolor{posgreen}+25.00 &  15 &  0.00 &   --   & --                         &   2 \\
Comprehensive Judgment    & 74.19 &  89.41 & \cellcolor{posgreen}+15.22 & 101 &  6.56 &  84.62 & \cellcolor{posgreen}+78.06 &  87 \\
Knowledge Application     & 63.64 &  87.50 & \cellcolor{posgreen}+23.86 &  86 & 16.67 &  33.33 & \cellcolor{posgreen}+16.67 &   7 \\
Statistical Methods       & 88.57 &  91.49 & \cellcolor{posgreen}+2.92  &  70 & 24.32 &  62.50 & \cellcolor{posgreen}+38.18 &  48 \\
\midrule
\textbf{Overall} & \textbf{78.22} & \textbf{88.58} & \textbf{+10.36} & \textbf{364} & \textbf{12.00} & \textbf{77.63} & \textbf{+65.63} & \textbf{304} \\
\bottomrule
\end{tabular}
\caption{Accuracy (\%) of GPT-OSS-20B by question type across English CFA and Chinese CPA examinations.}
\label{tab:gptoss20b_qtype}
\end{table*}

\paragraph{Detailed Results: GPT-OSS-120B.}

Table~\ref{tab:gptoss120b_topic} allows a direct scaling comparison at fixed training paradigm. A roughly 6$\times$ increase in total parameters yields less than three points of improvement in either English setting: \textit{Orig.}\ rises from 78.22\% to 80.39\% and \textit{Rev.}\ from 88.58\% to 89.10\%. On Chinese CPA, \textit{Orig.}\ accuracy actually declines from 12.00\% to 10.53\%. The Chinese dead-zone is therefore structural rather than capacity-bound. The Chinese \textit{Rev.}\ gain is larger for the 120B model (+74.05\%) than for the 20B (+65.63\%), indicating that the larger model commits to its default assumptions with greater confidence, not with greater caution.

\begin{table*}[t!]
\centering
\small
\begin{tabular}{@{}lrrrc|lrrrc@{}}
\toprule
\textbf{Question Topic (CFA)} & \textit{Orig.} & \textit{Rev.} & $\Delta$ & \textit{n} & \textbf{Question Topic (CPA)} & \textit{Orig.} & \textit{Rev.} & $\Delta$ & \textit{n} \\
\midrule
Corporate Finance              &  80.00 &  90.00 & \cellcolor{posgreen}+10.00 & 10  & Accounting        &  5.62 & 66.67 & \cellcolor{posgreen}+61.05 & 89 \\
Derivatives Investment         &  74.42 &  88.00 & \cellcolor{posgreen}+13.58 & 45  & Auditing          & 25.76 & 93.33 & \cellcolor{posgreen}+67.58 & 66 \\
Economics                      &  76.92 &  90.00 & \cellcolor{posgreen}+13.08 & 15  & Economic Law      &  6.67 & 91.89 & \cellcolor{posgreen}+85.23 & 60 \\
Equity Investment              &  85.96 &  88.46 & \cellcolor{posgreen}+2.50  & 59  & Tax Law           &  7.02 & 92.86 & \cellcolor{posgreen}+85.84 & 57 \\
Financial Statement Analysis   &  74.42 &  88.89 & \cellcolor{posgreen}+14.47 & 43  & Wealth Management &  6.25 & 75.00 & \cellcolor{posgreen}+68.75 & 32 \\
Fixed Income Investment        &  70.21 &  78.57 & \cellcolor{posgreen}+8.36  & 49  &                   &       &       &        &    \\
Other Investments              &  82.80 &  90.12 & \cellcolor{posgreen}+7.33  & 94  &                   &       &       &        &    \\
Portfolio Management           & 100.00 & 100.00 & +0.00                      &  4  &                   &       &       &        &    \\
Quantitative Analysis          &  89.36 &  91.11 & \cellcolor{posgreen}+1.75  & 47  &                   &       &       &        &    \\
\midrule
\textbf{Overall} & \textbf{80.39} & \textbf{89.10} & \textbf{+8.71} & \textbf{366} & \textbf{Overall} & \textbf{10.53} & \textbf{84.57} & \textbf{+74.05} & \textbf{304} \\
\bottomrule
\end{tabular}
\caption{Accuracy (\%) of GPT-OSS-120B by question topic. Left: English CFA (9 subjects). Right: Chinese CPA (5 subjects). \textit{n} denotes the number of original questions.}
\label{tab:gptoss120b_topic}
\end{table*}

Scaling compresses the dynamic range across English topics. \textit{Fixed Income Investment} rises from 63.41\% to 70.21\%, while \textit{Quantitative Analysis} is essentially unchanged. On Chinese CPA, the two rule-prescriptive topics most amenable to default substitution (\textit{Economic Law}, \textit{Tax Law}) exceed 91\% under \textit{Rev.}, while \textit{Wealth Management} still lags at 75.00\%. This asymmetry supports the main-paper claim that cross-product reasoning resists implicit assumption alignment more effectively than single-rule lookup does.

\begin{table*}[t!]
\centering
\small
\begin{tabular}{@{}lrrrr|rrrr@{}}
\toprule
 & \multicolumn{4}{c|}{\textbf{English CFA}} & \multicolumn{4}{c}{\textbf{Chinese CPA}} \\
\cmidrule(lr){2-5} \cmidrule(lr){6-9}
\textbf{Question Type} & \textit{Orig.} & \textit{Rev.} & $\Delta$ & \textit{n} & \textit{Orig.} & \textit{Rev.} & $\Delta$ & \textit{n} \\
\midrule
Conceptual Understanding  & 85.48 &  86.05 & \cellcolor{posgreen}+0.56  &  62 &  8.40 & 100.00 & \cellcolor{posgreen}+91.60 & 119 \\
Simple Calculation        & 93.33 &  87.71 & \cellcolor{negred}$-$5.62  &  30 &  4.88 &  69.81 & \cellcolor{posgreen}+64.93 &  41 \\
Complex Calculation       & 86.67 & 100.00 & \cellcolor{posgreen}+13.33 &  15 &  0.00 &    --  & --                         &   2 \\
Comprehensive Judgment    & 78.22 &  89.53 & \cellcolor{posgreen}+11.32 & 101 &  9.20 &  91.92 & \cellcolor{posgreen}+82.72 &  87 \\
Knowledge Application     & 63.64 & 100.00 & \cellcolor{posgreen}+36.36 &  86 & 28.57 &  66.67 & \cellcolor{posgreen}+38.10 &   7 \\
Statistical Methods       & 90.00 &  93.62 & \cellcolor{posgreen}+3.62  &  70 & 20.83 &  80.00 & \cellcolor{posgreen}+59.17 &  48 \\
\midrule
\textbf{Overall} & \textbf{80.39} & \textbf{89.10} & \textbf{+8.71} & \textbf{366} & \textbf{10.53} & \textbf{84.57} & \textbf{+74.05} & \textbf{304} \\
\bottomrule
\end{tabular}
\caption{Accuracy (\%) of GPT-OSS-120B by question type across English CFA and Chinese CPA examinations.}
\label{tab:gptoss120b_qtype}
\end{table*}

\paragraph{Robustness Analysis under Adversarial Masking.}

The NOTA-masked results for the GPT-OSS pair (Table~\ref{tab:answer-vs-reasoning-acc-gap}) provide the clearest evidence in our evaluation that parameter scaling alone does not resolve the structural failure we study.

GPT-OSS-20B exhibits a severe \textit{reasoning-selection gap}. \textit{Answer Accuracy} is 86.65\% on English and 67.43\% on Chinese, while \textit{Reasoning Accuracy} collapses to 12.26\% and 4.00\%. Qualitative inspection of the Harmony-format chains of thought confirms that the model seldom identifies the missing premise. Instead it commits to a specific numerical or regulatory default and proceeds as if the question were fully specified. The 20B variant also reports mean confidence $\mu=90.66$ on the English NOTA subset despite only 12.26\% reasoning accuracy, a clearer instance of the \textit{blind certainty} pattern first named for CFGPT2-7B (Appendix~\ref{app:cfgpt}).

Scaling to 120B raises \textit{Answer Accuracy} to 89.10\% on English and 84.57\% on Chinese. At face value this suggests substantial improvement. The \textit{Reasoning Accuracy} column disagrees. On English it rises modestly from 12.26\% to 20.71\%, still well below Fin-R1-7B's 42.50\% on the same subset. On Chinese it decreases from 4.00\% to 1.14\%. The answer--reasoning gap therefore widens with scale on Chinese, from 63.4 to 83.4 percentage points, and narrows only slightly on English, from 74.4 to 68.4 percentage points.

Confidence amplifies the concern. Mean confidence rises from 90.66 (20B) to 95.66 (120B) on English, and from 79.17 to 91.46 on Chinese. Both 120B values are the highest observed in our entire evaluation. The largest and most confident general-purpose reasoning model in our suite simultaneously shows the widest answer--reasoning gap. This is the central empirical argument against treating answer accuracy alone as a proxy for reliability in professional financial reasoning.



\subsection{DianJin-R1-32B Configuration}
\label{app:dianjin_r1}

\paragraph{Model Source.} \texttt{DianJin/DianJin-R1-32B}\footnote{\url{https://huggingface.co/DianJin/DianJin-R1-32B}}~\citep{DianJin-R1}

DianJin-R1-32B is a financial reasoning model developed by Alibaba Cloud, built on the Qwen2.5-32B base and trained with GRPO-style reinforcement learning on large-scale financial question-answering data. It is included in our evaluation as a larger and more recent finance-specialised reasoning model than Fin-R1-7B (Appendix~\ref{app:finr1}). Its inclusion tests whether the epistemic-caution gains observed at 7B scale under the same GRPO paradigm persist at 32B scale. We use a single NVIDIA H20 (96\,GB HBM3) with BF16 precision. At this precision the model occupies approximately 64\,GB and fits on a single GPU without tensor parallelism.

\paragraph{Chat Template.}
DianJin-R1-32B inherits the native Qwen-style chat template, applied through \texttt{tokenizer.apply\_chat\_template()} with \texttt{trust\_remote\_code=True}. We use the standard system/user message structure of our unified protocol (Appendix~\ref{app:prompts}):

\begin{figure}[h]
    \centering
    \tcbset{
  colframe=black!70,
  colback=gray!10,
  arc=2mm,
  boxrule=0.4pt,
  left=5pt,
  right=5pt,
  top=5pt,
  bottom=5pt}
    \begin{tcolorbox}
    \small
\begin{verbatim}
tokenizer = AutoTokenizer.from_pretrained(
    model_dir, trust_remote_code=True)
model = AutoModelForCausalLM.from_pretrained(
    model_dir,
    device_map="auto",
    torch_dtype="auto",
    trust_remote_code=True)
model.eval()

messages = [
    {"role": "system", "content": system_prompt},
    {"role": "user",   "content": user_prompt},
]
prompt = tokenizer.apply_chat_template(
    messages, tokenize=False,
    add_generation_prompt=True)

inputs = tokenizer(prompt, return_tensors="pt").
to(device)
out = model.generate(
    **inputs,
    max_new_tokens=1024,
    do_sample=False,
    repetition_penalty=1.05,
    eos_token_id=tokenizer.eos_token_id,
    pad_token_id=tokenizer.eos_token_id)
\end{verbatim}
    \end{tcolorbox}
    \caption{DianJin-R1-32B inference code (Qwen-style chat template, greedy decoding consistent with the rest of the suite).}
\end{figure}

\paragraph{Detailed Results.}

Table~\ref{tab:dianjin_topic} reflects the profile of a high-capacity finance-specialised reasoning model. English CFA \textit{Orig.}\ accuracy reaches 74.79\% and \textit{Rev.}\ reaches 89.26\% (+14.47\%), placing DianJin-R1-32B among the strongest open-weight finance models in our suite. Chinese CPA shows a much sharper contrast. \textit{Orig.}\ accuracy of 16.87\% is an order of magnitude above CFGPT2-7B's 2.96\% on the same benchmark. \textit{Rev.}\ accuracy rises to 93.79\% (+76.93\%), the highest Chinese \textit{Rev.}\ value in our entire evaluation. The size of the Chinese jump is consistent with the implicit assumption alignment pattern described in Section~\ref{sec:results}: a model trained on Chinese financial data aligns its regulatory and accounting defaults with the removed premises almost perfectly.

\begin{table*}[t!]
\centering
\small
\begin{tabular}{@{}lrrrc|lrrrc@{}}
\toprule
\textbf{Question Topic (CFA)} & \textit{Orig.} & \textit{Rev.} & $\Delta$ & \textit{n} & \textbf{Question Topic (CPA)} & \textit{Orig.} & \textit{Rev.} & $\Delta$ & \textit{n} \\
\midrule
Corporate Finance              &  80.00 &  90.00 & \cellcolor{posgreen}+10.00 & 10  & Accounting        & 11.29 & 85.71 & \cellcolor{posgreen}+74.42 & 89 \\
Derivatives Investment         &  63.64 &  96.00 & \cellcolor{posgreen}+32.36 & 45  & Auditing          & 42.11 & 96.55 & \cellcolor{posgreen}+54.45 & 66 \\
Economics                      &  75.00 &  90.00 & \cellcolor{posgreen}+15.00 & 15  & Economic Law      &  9.43 & 97.06 & \cellcolor{posgreen}+87.62 & 60 \\
Equity Investment              &  76.27 &  88.46 & \cellcolor{posgreen}+12.19 & 59  & Tax Law           &  9.80 & 97.37 & \cellcolor{posgreen}+87.56 & 57 \\
Financial Statement Analysis   &  76.74 &  88.89 & \cellcolor{posgreen}+12.14 & 43  & Wealth Management &  3.85 & 86.96 & \cellcolor{posgreen}+83.11 & 32 \\
Fixed Income Investment        &  51.11 &  78.57 & \cellcolor{posgreen}+27.46 & 49  &                   &       &       &        &    \\
Other Investments              &  80.65 &  88.89 & \cellcolor{posgreen}+8.24  & 94  &                   &       &       &        &    \\
Portfolio Management           & 100.00 & 100.00 & +0.00                      &  4  &                   &       &       &        &    \\
Quantitative Analysis          &  89.36 &  88.89 & \cellcolor{negred}$-$0.47  & 47  &                   &       &       &        &    \\
\midrule
\textbf{Overall} & \textbf{74.79} & \textbf{89.26} & \textbf{+14.47} & \textbf{366} & \textbf{Overall} & \textbf{16.87} & \textbf{93.79} & \textbf{+76.93} & \textbf{304} \\
\bottomrule
\end{tabular}
\caption{Accuracy (\%) of DianJin-R1-32B by question topic. Left: English CFA (9 subjects). Right: Chinese CPA (5 subjects). \textit{n} denotes the number of original questions.}
\label{tab:dianjin_topic}
\end{table*}

The Chinese \textit{Orig.}\ profile is non-trivial in its own right. \textit{Auditing} reaches 42.11\% and \textit{Accounting} 11.29\%, reflecting Chinese-financial knowledge acquired during continued pre-training. Under the \textit{Rev.}\ setting, the three rule-prescriptive topics (\textit{Economic Law}, \textit{Tax Law}, \textit{Auditing}) all exceed 96\%. The less rule-prescriptive \textit{Wealth Management} reaches 86.96\%, higher than for every other finance-specialised model in our evaluation. \textit{Rev.}\ peaks at 97.37\% on \textit{Tax Law}, just short of the ceiling.

\begin{table*}[t!]
\centering
\small
\begin{tabular}{@{}lrrrr|rrrr@{}}
\toprule
 & \multicolumn{4}{c|}{\textbf{English CFA}} & \multicolumn{4}{c}{\textbf{Chinese CPA}} \\
\cmidrule(lr){2-5} \cmidrule(lr){6-9}
\textbf{Question Type} & \textit{Orig.} & \textit{Rev.} & $\Delta$ & \textit{n} & \textit{Orig.} & \textit{Rev.} & $\Delta$ & \textit{n} \\
\midrule
Conceptual Understanding  & 80.65 &  88.10 & \cellcolor{posgreen}+7.45   &  62 & 16.22 & 100.00 & \cellcolor{posgreen}+83.78  & 119 \\
Simple Calculation        & 71.43 &  88.64 & \cellcolor{posgreen}+17.21  &  30 &  9.09 &  94.29 & \cellcolor{posgreen}+85.19  &  41 \\
Complex Calculation       & 64.29 & 100.00 & \cellcolor{posgreen}+35.71  &  15 &  0.00 &   --   & --                         &   2 \\
Comprehensive Judgment    & 84.16 &  89.53 & \cellcolor{posgreen}+5.38   & 101 & 16.90 &  92.31 & \cellcolor{posgreen}+75.41  &  87 \\
Knowledge Application     & 47.50 &  87.50 & \cellcolor{posgreen}+40.00  &  86 &  0.00 & 100.00 & \cellcolor{posgreen}+100.00 &   7 \\
Statistical Methods       & 90.00 &  91.49 & \cellcolor{posgreen}+1.49   &  70 & 25.64 & 100.00 & \cellcolor{posgreen}+74.36  &  48 \\
\midrule
\textbf{Overall} & \textbf{74.79} & \textbf{89.26} & \textbf{+14.47} & \textbf{364} & \textbf{16.87} & \textbf{93.79} & \textbf{+76.93} & \textbf{304} \\
\bottomrule
\end{tabular}
\caption{Accuracy (\%) of DianJin-R1-32B by question type across English CFA and Chinese CPA examinations.}
\label{tab:dianjin_qtype}
\end{table*}

\paragraph{Robustness Analysis under Adversarial Masking.}

DianJin-R1-32B exhibits the most pronounced English-Chinese asymmetry among all evaluated models. On the English NOTA subset \textit{Answer Accuracy} reaches 88.28\% and \textit{Reasoning Accuracy} reaches 29.16\%, a 59.1-point gap. This gap is substantially narrower than GPT-OSS-120B's 68.4-point gap on the same subset and is broadly consistent with the behaviour of the smaller Fin-R1-7B (Appendix~\ref{app:finr1}). GRPO-style reasoning RL, when transferred to a larger 32B Qwen2.5 base, therefore preserves a non-trivial share of English condition-detection capability.

The Chinese picture is sharply different. \textit{Answer Accuracy} remains high at 77.71\%, in line with the 93.79\% Chinese \textit{Rev.}\ accuracy reported above. \textit{Reasoning Accuracy} drops to 5.71\%, a 72.0-point gap. Qualitative inspection shows that Chinese explanations rarely articulate that a required premise is missing. The model commits to a specific regulatory or accounting default and returns the option consistent with that default. The 93.79\% Chinese \textit{Rev.}\ value therefore reflects successful \emph{alignment} of training-time defaults with the benchmark's removed premises, not recognition of under-specification.

Confidence is moderate by the standards of the extended suite. Mean confidence is 93.71 on English and 77.80 on Chinese, both below GPT-OSS-120B (95.66 and 91.46) and broadly comparable to Fin-R1-7B. DianJin-R1-32B thus exhibits the signature GRPO-trained profile: high answer accuracy, measurable English reasoning consistency relative to non-RL baselines, and moderate confidence. It does not escape the core Chinese-CPA failure mode that persists across every model family in our evaluation.

\end{document}